\documentclass{article}
\usepackage[version=3]{mhchem}
\usepackage[utf8]{inputenc}
\usepackage[dvipsnames]{xcolor}
\usepackage{authblk}
\usepackage{amsmath}
\usepackage{graphicx}
\usepackage{geometry}
\usepackage{hyperref}
\usepackage{color, colortbl}
\definecolor{Gray}{gray}{0.95}

\begin{document}

\title{Enhancing metallicity and basal plane reactivity of 2D materials via self-intercalation}
\author[1,*]{Stefano Americo}
\author[1]{Sahar Pakdel}
\author[1]{Kristian Sommer Thygesen}
\affil[1]{Computational Atomic-scale Materials Design (CAMD), Department of Physics, Technical University of Denmark, 2800 Kgs. Lyngby Denmark}
\affil[*]{Email: steame@dtu.dk}
\maketitle


\section{Abstract}
Intercalation (ic) of metal atoms into the van der Waals (vdW) gap of layered materials constitutes a facile strategy to create new materials whose properties can be tuned via the concentration of the intercalated atoms. Here we perform systematic density functional theory calculations to explore various properties of an emergent class of crystalline 2D materials (ic-2D materials) comprising vdW homobilayers with native metal atoms on a sublattice of intercalation sites. From an initial set of 1348 ic-2D materials, generated from 77 vdW homobilayers, we find 95 structures with good thermodynamic stability (formation energy within 200 meV/atom of the convex hull). A significant fraction of the semiconducting host materials are found to undergo an insulator to metal transition upon self-intercalation with only PdS$_2$, PdSe$_2$, and GeS$_2$ maintaining a finite electronic gap. In five cases, self-intercalation introduces magnetism. In general, self-intercalation is found to promote metallicity and enhance the chemical reactivity on the basal plane. Based on the calculated H binding energy we find that self-intercalated \ce{SnS2} and \ce{Hf3Te2} are promising candidates for hydrogen evolution catalysis. All the stable ic-2D structures and their calculated properties can be explored in the open C2DB database.

\section{Introduction}
Atomically thin two-dimensional (2D) materials exhibit unique physical properties that could form the basis for future generations of ultra-compact devices with novel functionalities or architecture designs.\cite{das2021transistors,cao20212d,ahn20202d,wang2023towards} Moreover, the extreme surface area-to-volume ratio makes 2D materials natural candidates for chemical sensors\cite{lee2019chemical,anichini2018chemical} and (electro)catalysts\cite{cummins2016efficient,pandey2017two,pandey2015two,li20192d,zhang2018single,mondal20222d} -- in particular if their basal planes could be activated.\cite{han2018activation,yang2019ultrahigh} So far, most research has focused on 2D monolayers and their stacked homo- or heterostructures.\cite{jariwala2014emerging,briggs2019roadmap} While stacking can be used to modulate the electronic and optical properties of 2D materials\cite{peimyoo2021electrical,tran2019evidence,winther2017band}, the effects are relatively limited due to the weakness of the interlayer van der Waals (vdW) interactions. An alternative functionalisation strategy consists in intercalating foreign or native atoms into the vdW gap of a multi-layer structure. In general, this scheme can produce significant changes to the materials properties thanks to the strong covalent bonds formed between the host 2D layers and the intercalated atoms.\cite{rajapakse2021intercalation,wu2023electrostatic,zhang2023mechanism} In particular, it is expected that intercalation can enhance interlayer interactions, induce doping of the 2D materials, strongly modify their electronic structure and chemical reactivity, or even convert them into distinctly different structural phases. 

Self-intercalation, i.e. intercalation of native metal atoms, has recently emerged as an effective technique to produce new types of covalently bonded, crystalline 2D materials (referred to as ic-2D materials) with stoichiometry dependent properties\cite{zhao2020engineering} and controllable thickness.
This approach is interesting for several reasons: (1) Intercalation in few-layer structures, in particular bilayers, provides an effective way to expand the family of atomically thin 2D materials. (2) The properties of ic-2D materials can be tuned by varying the concentration of intercalated atoms, i.e. the stoichiometry. (3) ic-2D materials with different stoichiometries, including the parent vdW host structure, belong to the same compositional space yet may exhibit complementary properties. In particular, they share the same chemical elements and have similar lattice constants, but may be metallic or insulating, magnetic or non-magnetic, and show different chemical properties depending on the stoichiometry. Such compatible/complementary material sets are ideal for device construction.     

Zhao \emph{et al.} produced a range of ic-2D materials, in particular TaS$_2$ and TaSe$_2$ with various concentrations of intercalated Ta, using both molecular beam epitaxy (MBE) and chemical vapour deposition (CVD). \cite{zhao2020engineering} In these experiments, the stoichiometry was controlled during growth by tuning the relative chemical potentials of Ta and S/Se, and the resulting ic-2D materials showed stoichiometry-dependent crystal structures and magnetism. 
There have been several reports on CVD and MBE grown few-layer Cr$_2$Te$_3$, which is a self-intercalated version of the vdW crystal CrTe$_2$, showing thickness dependent conduction behaviour (from n- to p-type)\cite{cui2020controlled} and tunable ferromagnetism down to the limit of intercalated bilayers\cite{li2019molecular,wen2020tunable,lasek2022van}. It has been shown that self-intercalation compounds competes with and often dominates over multilayer vdW crystals during direct MBE growth of the early transition metal (Ti, V, Cr) ditellurides.\cite{lasek2020molecular} 
Yang \emph{et al.} showed that Nb intercalation can dramatically increase the catalytic activity of the vdW crystal NbS$_2$ for the hydrogen evolution reaction\cite{yang2019ultrahigh}.  

In addition to these experimental studies, density functional theory calculations have shown that self-intercalation can modulate and enhance the interlayer exchange interaction and the magnetic anisotropy in intrinsically magnetic vdW materials such as CrTe$_2$\cite{li2021magnetic}, FeCl$_2$\cite{li2022tailoring}, and CrI$_3$\cite{guo2020enhanced}.

In this work, we perform a systematic exploration of ic-2D materials and their stoichiometry dependent properties using density functional theory (DFT) calculations. Starting from 77 binary vdW bilayers, we insert native metal atoms in the vdW gap in various concentrations and determine the stability of the resulting crystal structures. For the most stable compounds, we compute a range of properties including the mechanical stiffness, electronic band structures, magnetic moments, and basal plane reactivity (defined from the adsorption energy of H). Out of 23 semiconducting vdW bilayers, 20 undergo an insulator-to-metal transition upon self-intercalation. This transition is clearly correlated with an increase in the H binding energy. Our results show that intercalation is an efficient means to activate the basal plane of 2D materials. Specifically, we identify seven ic-2D materials (involving five different host systems) as promising electrocatalysts for the hydrogen evolution reaction.

\section{Results and discussion}    
We start by describing the workflow used to set up the ic-2D structures. In the subsequent sections, we discuss the thermodynamic stability of the resulting ic-2D structure and their agreement with available experimental data before we explore their electronic properties and potential as electrocatalysts for the hydrogen evolution reaction.  

\paragraph{Intercalation workflow}

 \begin{figure}[ht]
        \includegraphics[width=\textwidth]{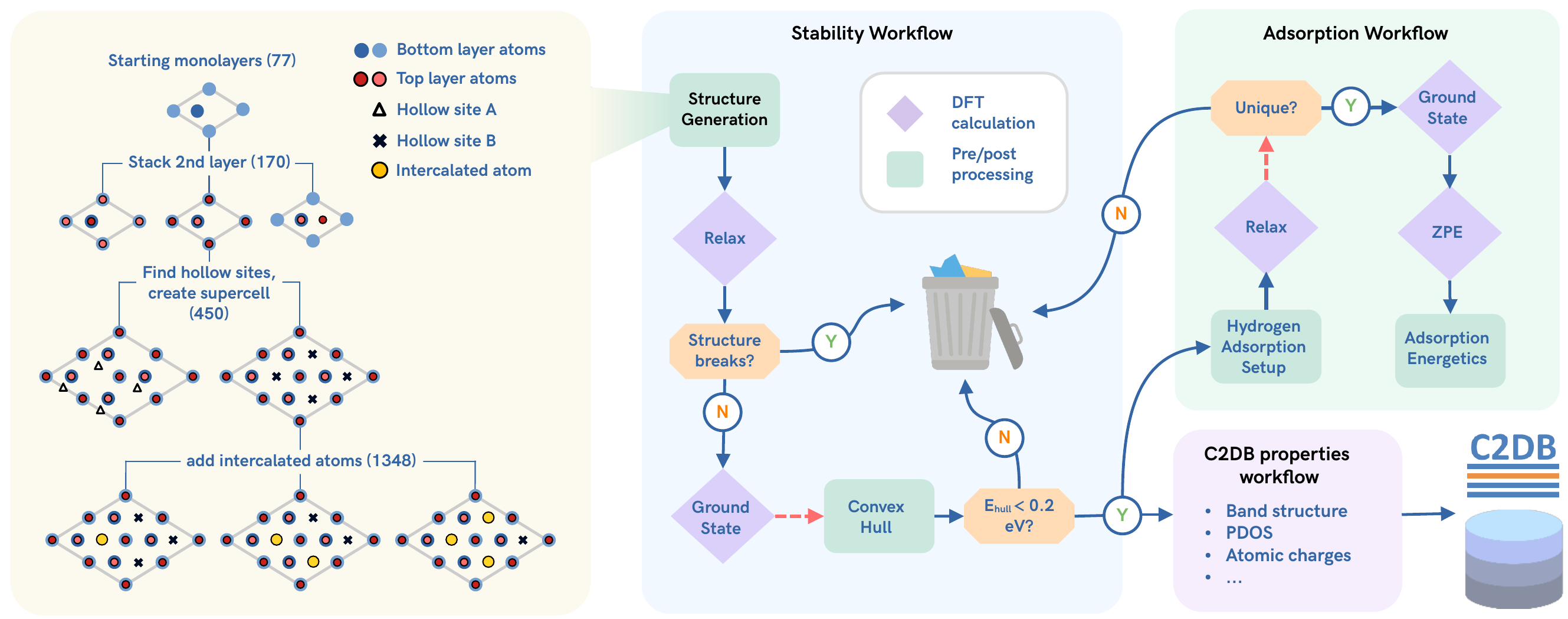}
        \caption{
            Workflow for calculation of the ic-2D properties, including 
            a schematic representation of the structure generation procedure.
            The numbers in parentheses represent the number of structures generated
            at each step.
            Red dashed arrow represent bottlenecks in the workflow,
            where all the previous steps have to be completed for the related 
            ic-2D structures before proceeding to the following steps.
        }
        \label{fig:WF-structgen}
    \end{figure}

All the ic-2D bilayer structures are generated 
starting from a set of parent monolayers (ML) selected from the Computational 2D Material Database (C2DB)\cite{haastrup2018computational,gjerding2021recent} according to the 
following criteria:

a) At most two different chemical species (A$_x$B$_y$).

b) Good thermodynamic stability. Specifically, the energy above the convex hull of the ML ($\Delta H_{\mathrm{hull}}$) is below 100 meV/atom.

c) Dynamical stability, i.e. all phonon energies of the ML should be real and positive.

d) Maximum six atoms per unit cell in the ML. This condition limits the complexity of the resulting ic-2D structures and the computational demands. 

e) The material should be experimentally known in few-layer or bulk form. In practice this means that all the materials are reported in either the ICSD\cite{bergerhoff1987crystallographic} and/or the COD\cite{gravzulis2012crystallography} crystal structure databases and/or
can have been explored experimentally and reported in the literature.

f) The ML does not contain rare\cite{taylor1964abundance} or toxic elements.
The blacklisted elements are 
Cd, Pb, Hg, Pt, Au, Os, Ru, Ir, Rh, Re, Kr, Xe.
All materials are thus ensured to be obtainable from earth-abundant elements and suitable for 
sustainable applications.

These preliminary selection criteria result in a set of 77 MLs. The ic-2D bilayer generation procedure follows the 
scheme illustrated in Fig. \ref{fig:WF-structgen} (left). For each of the MLs, 
we obtain a set of vdW homobilayers corresponding to the most stable stacking 
configurations as defined by Pakdel\cite{pakdel2023emergent}. For each stacking,
we automatically identify all the hollow sites in the vdW gap region that are inequivalent by symmetry (see Methods section for a description of the software framework in use). We then determine a suitable supercell containing at least three sites for each hollow site type, typically resulting in a ($\sqrt{3} \times \sqrt{3}$)R30 or a 2$\times$2 supercell depending on the host bilayer cell symmetry.
For each hollow site type, we then generate ic-2D bilayers with different concentrations
of intercalated atoms. 

Defining the ic-fraction as the number of intercalated atoms per primitive unit cell of the bilayer host (for instance a 2$\times$2 supercell with one intercalated atom will have an ic-fraction of 0.25), 
we obtain fractions in the 0.25 - 1.0 range, which are comparable with the ones typically realized in experiments\cite{zhao2020engineering,yang2019ultrahigh,lasek2020molecular}. In the following, this fraction will sometimes be referred to as a percentage or ic-concentration.
We note that the relatively high concentrations considered implies that the ic-atoms should not be regarded as
point defects/impurities. Rather, they form a periodic sublattice producing a completely new crystalline phase with well-defined stoichiometry.

Following this procedure, the initial 77 MLs result in a total of 1348 ic-2D bilayers.
Due to the large number of structures, we developed a high-throughput workflow to manage the DFT calculations.
As illustrated in Fig. \ref{fig:WF-structgen}, the workflow consists of the three sub-sections described below: 

\textit{Stability workflow.} First, the ic-2D atomic structures are relaxed using the PBE exchange-correlation functional. Structures that undergo major rearrangements during relaxation or present no significant chemical bonding between the the intercalated atoms and the 2D layers, are discarded. The remaining ic-2D structures represent covalently bonded 2D materials justifying the use of the PBE functional. In particular, we find that the use of the D3 correction to account for vdW interactions has a negligible effect on on the structures (see Fig. S1).   
The workflow then groups the ic-2D bilayers according to the parent MLs, and calculates the energy above the convex hull using the MLs from C2DB and most stable binary and ternary bulk compounds   
from the OQMD database\cite{saal2013materials} as reference phases. 
For each ic-concentration, the workflow selects the stacking/hollow site combination that yields the lowest formation energy, 
and keeps the structure if its energy above the convex hull fulfills $\Delta H_{\mathrm{hull}}<0.2$ eV/atom. Since we find that the electronic properties (\textit{e.g.} the band gap) are not very sensitive to the precise intercalation site or the stacking order, we only consider the most stable ic-2D material for each ic-concentration. 
The selected ic-2D bilayers are then passed to the "Adsorption" and "Properties" workflows (right panels in Fig. \ref{fig:WF-structgen}).

\textit{Adsorption workflow.} 
Different configurations are generated by adsorbing a hydrogen atom at each of the inequivalent adsorption 
sites at the surface of the ic-2D structure. Since hydrogen typically shows weak adsorbate-adsorbate interactions\cite{gossenberger2020sulfate}, even at our relatively high hydrogen coverage (20-30\%) the adsorbed atoms can be
considered as isolated.
After performing a structural optimization, the workflow filters out duplicate structures (initially different structures may become identical after relaxation). 
In order to reduce the overall computational cost, the zero point energy (ZPE) of the adsorbed hydrogen is calculated only for 
the lowest-energy configuration and used for all the remaining configurations when evaluating the adsorption free energies. 

\textit{Properties workflow.}  The workflow evaluates basic properties of the ic-2D materials including the stiffness tensor,
the electronic band structure, the projected density of states, Fermi surface, Bader charges, and magnetic moments. This workflow is a part of the full property workflow used to characterise 2D materials in the C2DB.

\paragraph{Thermodynamic stability}

\begin{table}[ht]
    \centering
    \resizebox{0.75\textwidth}{!}{
\begin{tabular}{cccccc}
\hline \hline
               Parent ML  &  ic-fraction  &  $\Delta H_{f}$ (eV/atom)  &  $\Delta H_{\mathrm{hull}}$ (meV/atom)  &  $E_{\mathrm{gap}}$ (eV) & $\mu_{\mathrm{mag}}$  \\
\hline
\rowcolor{Gray}
  \ce{Bi2Te3}  &    0.33     &     -0.18     &            45.2            &               0               &              0              \\
  \ce{CrTe2}   &     0.25     &     -0.23     &            22.8             &               0               &            23.70            \\
               &      0.5      &     -0.23     &            39.7            &               0               &            28.74            \\
\rowcolor{Gray}
   \ce{GaTe}   &    0.33     &     -0.32     &            40.3            &               0               &              0              \\
 \ce{HfS2}(T)  &    0.33     &     -1.57     &           -39.2            &               0               &              0              \\
               &    0.67     &     -1.52     &           -34.2            &               0               &              0              \\
               &      1      &     -1.4      &            29.8            &               0               &              0              \\
\rowcolor{Gray}
 \ce{HfSe2}(T) &    0.33     &     -1.31     &            4.1             &               0               &              0              \\
 \ce{HfTe2}(T) &    0.33     &     -0.8      &            -2.1            &               0               &              0              \\
               &    0.67     &     -0.73     &            45.7            &               0               &              0              \\
\rowcolor{Gray}
   \ce{InSe}   &    0.33     &     -0.49     &             24             &               0               &              0              \\
 \ce{NbS2}(H)  &    0.33     &     -1.17     &           -51.6            &               0               &              0              \\
\rowcolor{Gray}
 \ce{NbS2}(T)  &    0.33     &     -1.1      &            23.6            &               0               &              0              \\
\rowcolor{Gray}
               &    0.67     &     -1.17     &           -12.1            &               0               &              0              \\
 \ce{NbSe2}(H) &    0.33     &     -0.9      &           -10.3            &               0               &              0              \\
               &    0.67     &     -0.88     &            22.7            &               0               &              0              \\
\rowcolor{Gray}
 \ce{NbSe2}(T) &    0.33     &     -0.85     &            41.8            &               0               &              0              \\
   \ce{PdS2}   &    0.25     &     -0.27     &            30.5            &             0.75              &              0              \\
\rowcolor{Gray}
  \ce{PdSe2}   &    0.25     &     -0.28     &            13.1            &             0.79              &              0              \\
\rowcolor{Gray}
               &     0.5     &     -0.27     &            32.8            &             0.71              &              0              \\
 \ce{SnS2}(T)  &    0.33     &     -0.38     &            35.5            &               0               &              0              \\
\rowcolor{Gray}
 \ce{SnSe2}(T) &    0.33     &     -0.36     &            9.8             &               0               &              0              \\
\rowcolor{Gray}
               &    0.67     &     -0.36     &            17.5            &               0               &              0              \\
\rowcolor{Gray}
               &      1      &     -0.38     &            16.8            &               0               &              0              \\
 \ce{TaS2}(H)  &    0.33     &     -1.13     &           -75.3            &               0               &              0              \\
               &    0.67     &     -1.09     &           -70.2            &               0               &              0              \\
               &      1      &     -1.01     &           -17.8            &               0               &              0              \\
\rowcolor{Gray}
 \ce{TaS2}(T)  &    0.33     &     -1.05     &            2.3             &               0               &              0              \\
\rowcolor{Gray}
               &    0.67     &     -0.99     &            31.4            &               0               &              0              \\
 \ce{TaSe2}(T) &    0.33     &     -0.77     &            34.5            &               0               &              0              \\
\rowcolor{Gray}
 \ce{TiTe2}(T) &    0.33     &     -0.72     &            -6.1            &               0               &              0              \\
\rowcolor{Gray}
               &    0.67     &     -0.69     &            28.7            &               0               &              0              \\
  \ce{VS2}(H)  &    0.33     &     -0.96     &            -26             &               0               &            1.51             \\
               &    0.67     &     -0.96     &            23.8            &               0               &              0              \\
\rowcolor{Gray}
   \ce{W2N3}   &    0.33     &     -0.32     &            -135            &               0               &              0              \\
\rowcolor{Gray}
               &    0.67     &     -0.31     &           -121.1           &               0               &              0              \\
\rowcolor{Gray}
               &      1      &     -0.2      &           -15.6            &               0               &              0              \\
 \ce{ZrSe2}(T) &    0.33     &     -1.38     &           -60.4            &               0               &              0              \\
               &    0.67     &     -1.35     &           -65.7            &               0               &              0              \\
               &      1      &     -1.3      &           -49.8            &               0               &              0              \\
\hline
\end{tabular}
}

    \caption{
        List of ic-2D materials with $\Delta H_{\mathrm{hull}} < 50$ meV/atom. For each material, we report the chemical formula of the
        parent monolayer, the intercalated fraction, the heat of formation, the energy above the convex hull, 
        the band gap, and the magnetic moment.
    }
    \label{tab:summary}
\end{table}

\begin{figure}[ht]
    \includegraphics[width=\textwidth]{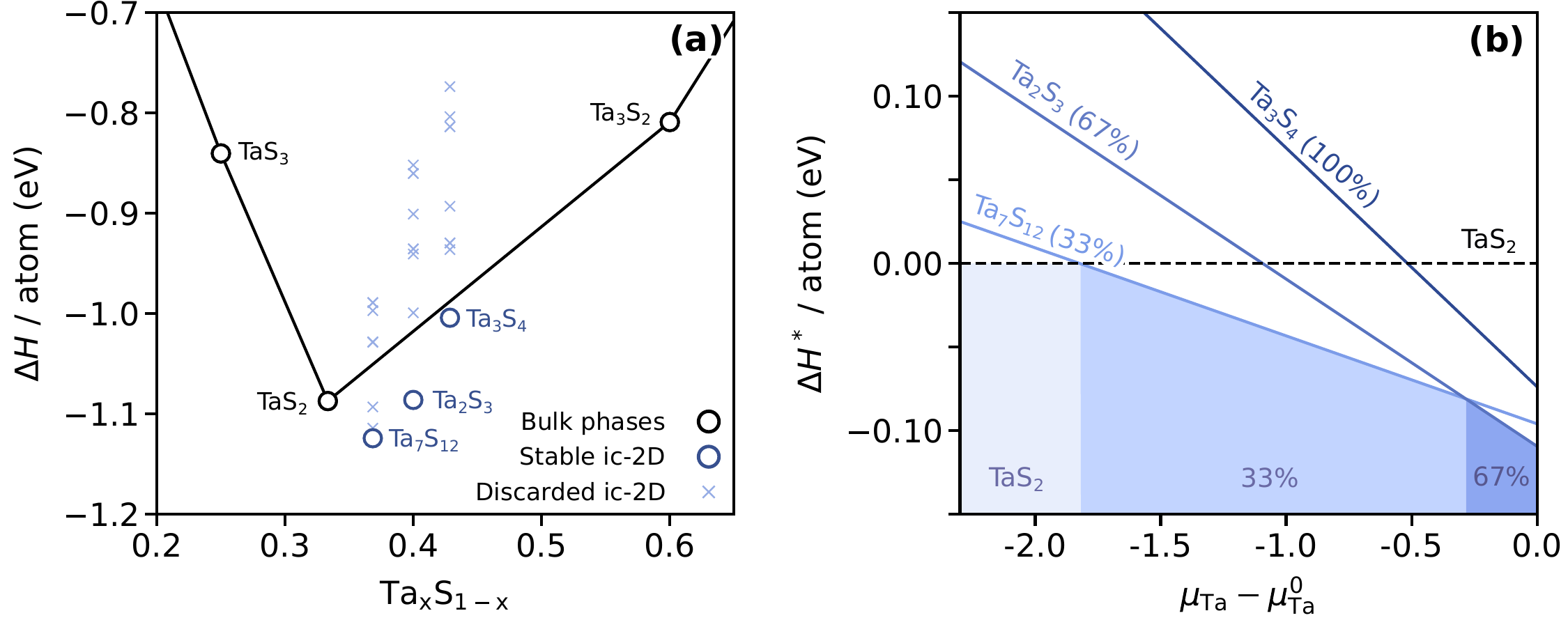}
    \caption{
        a) Ta-S convex hull including the stable self-intercalated phases (blue circles).
        The other tested configurations with higher formation energies 
        are represented by blue crosses.
        b) Formation energy of the ic-2D phases as a function of the Ta chemical potential 
        ($\mu_{\mathrm{Ta}}$) relative to the standard state. The asterisk in $\Delta H^*$
        indicates that here $\mu_{\mathrm{Ta}}$ is treated as a free variable, 
        while the S chemical potential $\mu_{\mathrm{S}}$ is constrained such that \ce{TaS2}
        formation is always at equilibrium, \textit{i.e.}
        $\mu_{\mathrm{Ta}} + 2~\mu_{\mathrm{S}} = E(\ce{TaS2})$.
        The ic-concentrations are shown in parentheses. The regions corresponding to
        different stable phases have been colored in blue.
    }
    \label{fig:convehull-TaS2}
\end{figure}
    
As previously described, the initial part of the workflow evaluates the thermodynamic stability of the candidate ic-2D structures. For each of the 77 host materials, the workflow collects the set of structures with identical stoichiometry (equivalently, the same ic-concentration). For each subset of structure, only the one with the lowest formation energy is kept. This results in 117 ic-2D structures that also satisfy our stability criterion of $\Delta H_{\mathrm{hull}}<0.2$ eV/atom. 

We further exclude the ic-2D structures where the distance between the intercalation layer and any of the layers of the host system is larger
than 3 \AA. This indicates that the intercalation layer does not provide sufficient binding energy,
resulting in a weakly interacting three-layer system rather than a new, covalently bonded phase.

Our final set of materials consists of the 95 ic-2D materials with $\Delta H_{\mathrm{hull}}<0.2$ eV/atom listed in Table S1-2. Among these, we identify six different structural archetypes (shown in Fig. S2) based on the structure and stoichiometry of the host system.
The subset of 40 ic-2D with $\Delta H_{\mathrm{hull}}<50$ meV/atom is shown in Table \ref{tab:summary}.

Out of the initial 77 host materials, 39 have at least one ic-2D structure with $\Delta H_{\mathrm{hull}}<0.2$ eV/atom.
As a general trend, the thermodynamic stability tends to decrease (i.e. the formation energy increases) for higher ic-concentrations. Several materials, however, satisfy our stability
criterion even at high ic-concentrations. For instance, Fig. \ref{fig:convehull-TaS2}a shows the calculated convex hull 
for the ic-2D structures obtained from \ce{TaS2} -- one of
the materials for which self-intercalation in the same concentration range has been realized experimentally.\cite{zhao2020engineering}
Fig. \ref{fig:convehull-TaS2}b shows the relative stability of the ic-2D
phases as a function of the Ta chemical potential ($\mu_{\ce{Ta}}$), highlighting how, at progressively higher $\mu_{\ce{Ta}}$ (Ta-rich environment), the pristine bilayer is replaced by its 33\%
and 67\% intercalated phases.

We also find that the formation energy of the ic-2D tends to assume more negative values for the ic-2D where the ratio between the atomic radii of the metallic atom and the non-metallic atom is relatively large. This general trend is shown in Fig. S3.

\paragraph{Comparison with experiments}
Self-intercalation has been experimentally demonstrated for a number of the vdW materials considered in this work, namely 
\ce{TaS2}\cite{zhao2020engineering},
\ce{TaSe2}\cite{zhao2020engineering}, \ce{VS2}\cite{zhao2020engineering,yang2019vanadium,zhang2021deciphering}, \ce{NbS2}\cite{yang2019ultrahigh}, \ce{CrTe2}\cite{lasek2020molecular}, \ce{VTe2}\cite{lasek2020molecular} and \ce{TiTe2}\cite{lasek2020molecular}. It is highly encouraging that for all these materials, the most stable ic-2D structure found by our workflow for all considered ic-concentrations, satisfy our general stability requirement of $\Delta H_{\mathrm{hull}}<0.2$ eV/atom. In fact, for most of the materials we find an ic-2D structure with negative $\Delta H_{\mathrm{hull}}$ and for all the materials we always find at least one ic-concentration satisfying $\Delta H_{\mathrm{hull}}<50$ meV/atom. Below we provide a more detailed discussion for each of the materials.  

ic-\ce{TaS2} has been synthesised by CVD with ic-concentrations in the 25\% to 100\% range\cite{zhao2020engineering}. In good agreement, we find the 33\%, 67\% and 100\% ic-2D structures to be thermodynamically stable ($\Delta H_{\mathrm{hull}}<0$), thus confirming that \ce{TaS2} constitutes a robust host system for self-intercalation over a broad concentration range.
For the 33\% ic-2D structure, the calculated in-plane lattice constant of 5.7 \AA~compares well with the experimental value of 5.8 \AA. The calculated distance between metal atoms belonging to the two \ce{TaS2} layers is 6.0 \AA~while the experimental value is slightly larger, namely 6.6 \AA.

ic-2D structures have also been experimentally realised for \ce{TaSe2} in the same concentration range\cite{zhao2020engineering}, although in this work we find lower stability towards intercalation as compared to \ce{TaS2}. 
The 33\%, 67\% and 100\% phases all satisfy our $\Delta H_{\mathrm{hull}}$ stability criterion, although the structures are predicted to be metastable relative to the pristine bilayer and Ta atoms in the solid phase. Specifically, the energy above the convex hull ($\Delta H_{\mathrm{hull}}$) is 34, 105, and 176 meV, respectively.
The calculated in-plane lattice constant is 5.9~\AA for both the 67\% and 100\% ic-2D structures, which is underestimated with respect to the experimental values (6.6 and 6.2 \AA, respectively).

Self-intercalated vanadium disulphide \ce{VS2} has been successfully synthesized at concentrations of 20-25\%\cite{zhao2020engineering,yang2019vanadium,zhang2021deciphering}. In this work, we find the 33\%-ic configuration to be thermodynamically stable and the 66\%-ic as metastable, lying 24 meV above the convex hull.

\ce{TiTe2} has been self-intercalated experimentally at 50\% ic-concentration\cite{lasek2020molecular}. All three ic-2D structures generated in this work satisfy our stability criterion. The 66\% and 100\% ic-concentrations being slightly metastable (29 and 74 meV above the hull, respectively) and the 33\% being thermodynamically stable.

For \ce{CrTe2}, a 50\% ic-2D structure has been synthesised\cite{lasek2020molecular}. We predict the 25\% and 50\% ic-2D structures to be slightly meta-stable with $\Delta H_{\mathrm{hull}}$ values of 30 and 40 meV, respectively. The calculated $d_{\mathrm{m-m}}$ of 6.3 \AA~compares well with the experimental value of 5.9 eV.

Finally, multilayer \ce{NbS2} has been self-intercalated experimentally\cite{yang2019ultrahigh} in a form, which we believe corresponds to the 33\% ic-concentration based on the provided stoichiometry and STEM images. We find the 33\% and 67\% ic-2D structures to be thermodynamically stable while the 100\% ic-2D structure is predicted to be metastable with $\Delta H_{\mathrm{hull}}$ of 59 meV. The calculated vertical distance between Nb atoms of the two \ce{NbS2} layers is 6.5 \AA~in good agreement with the experimental value of 6.8 \AA.

\paragraph{Electronic properties}

 \begin{figure}
        \includegraphics[width=\textwidth]{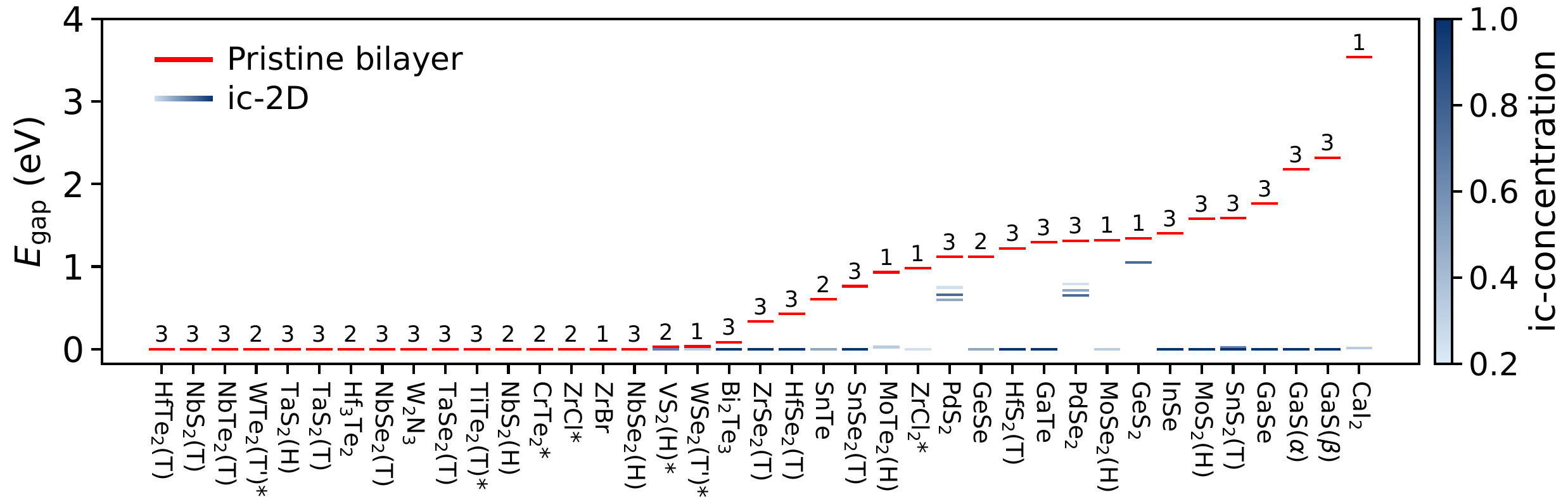}
        \caption{
            Band gap ($E_g$) of the pristine bilayer host systems (in red)
            and their ic-2D counterparts (in shades of blue).
            The x-axis displays the monolayer unit formulas and, 
            in parenthesis, for some materials the crystalline phase of the monolayer.
            Sytems with the same stoichiometry are labelled with greek
            letters when a reported crystalline phase is missing.
            An asterisk is appended to the label for systems with at least one magnetic ic-2D configuration.
            The ic-2D are reported in shades of blue according to the ic-concentration, following the color map on the side.
            Only ic-2D with $\Delta H_{hull} \leq 0.2$ eV/atom are considered. 
            For each starting bilayer, the number of ic-2D configurations that satisfy this condition is reported above the pristine bilayer line.
        }
        \label{fig:gapchange}
    \end{figure}
    
 \begin{figure}
        \includegraphics[width=\textwidth]{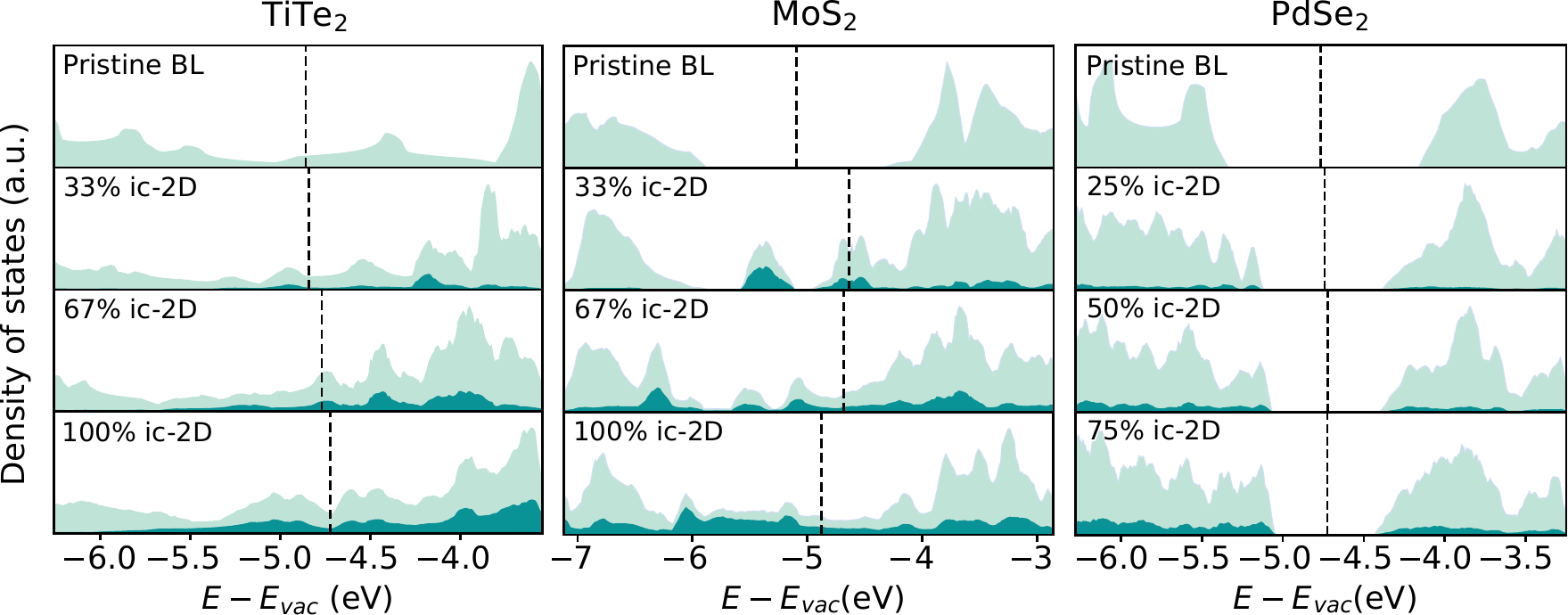}
        \caption{
            Density of states of ic-2D systems obtained from 
            \ce{TiTe2}, \ce{MoS2} and \ce{PdSe2}.
            For each, the top graph represents the total density of states
            of the corresponding pristine bilayer.
            The total density of states is shown in light blue, while the projections
            on the self-intercalated atoms are shown in darker blue.
            Vertical dashed lines represent the position of the
            Fermi level. Energies on the x-axis are relative to the vacuum
            level $E_{vac}$
        }
        \label{fig:PDOS}
    \end{figure}
    
 \begin{figure}
        \includegraphics[width=0.95\textwidth]{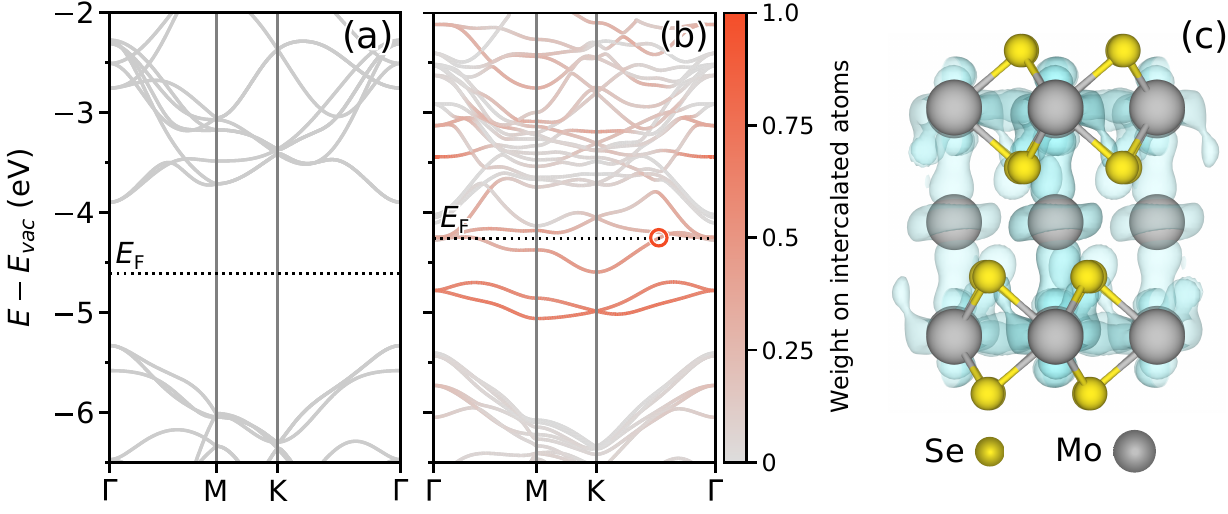}
        \caption{
            a) Band structure of a pristine \ce{MoSe2} monolayer in a 
            ($\sqrt{3} \times \sqrt{3}$)R30 supercell. Energies are relative to the 
            vacuum level, the Fermi energy is represented by a dashed line.
            b) Band structure of 33\% self-intercalated \ce{MoSe2},
            in the same supercell as for panel a). The color scale shows the projections of
            the eigenstates on the intercalated atoms.
            c) Real space distribution of the states highlighted by the red circle in panel b).
        }
        \label{fig:BS-WFs}
    \end{figure}
    
The primary aim of this work is to investigate the viability of self-intercalation as a strategy to functionalize 2D materials. In order to assess how self-intercalation affects the intrinsic properties of the bilayer we have calculated the electronic band structures of the 95 most stable ic-2D structures and compared them to those of the pristine vdW bilayers. 

Fig. \ref{fig:gapchange} shows the electronic band gap of the 95 ic-2D structures (blue lines) and the pristine bilayers (red lines). The ic-concentration is indicated by the blue color code. Materials for which a magnetic ground state is found for at least one ic-concentration, are indicated by an asterisk. It can immediately be  observed that self-intercalation always preserves the metallic nature when it is already present in the pristine bilayer.
When the host bilayer exhibits a finite band gap, this is either fully eliminated or significantly reduced upon self-intercalation. The enhanced metallicity observed in ic-2D obtained from semiconducting host systems agrees with the findings by Coelho \textit{et al.} on self-intercalated \ce{MoSe2} and \ce{MoTe2}\cite{coelho2018post}.

Fig. \ref{fig:PDOS} illustrates the evolution of the density of states (DOS) as a function of ic-fraction for three representative systems, namely \ce{TiTe2}, \ce{MoS2} and \ce{PdSe2}. These systems represent the three distinct cases of: (i) a metallic bilayer remaining metallic, (ii) a gapped bilayer evolving into a metal, and (iii) a semiconducting bilayer that remains gapped.   
Although intercalation does not change the metallic nature of \ce{TiTe2}, it significantly increases the DOS at the
Fermi level. Thus even in cases of metals, self-intercalation enhances the metallicity. As illustrated by the case of \ce{MoS2} (middle panel of Fig. \ref{fig:PDOS}) the transition from an insulator to a metal takes place already at relatively low ic-concentrations and occurs due to formation of new (metallic) states inside the band gap. 

As an example, Fig. \ref{fig:BS-WFs}b shows the band structure of the 33\%-intercalated \ce{MoSe2} (H-phase), which has a 1.3 eV PBE band gap in its ML form. Bands with larger weight on the ic-atoms appear more red. Compared to the ML band structure, shown in Fig. \ref{fig:BS-WFs}a, it can be seen how bands with substantial weight on the intercalated atoms appear in the gap region, now crossing the Fermi level.
Due to the covalent nature of the bonding between the intercalated atoms and the 2D host layers, such states are never fully localized on the ic-atoms and typically carry significant contributions from the surrounding atomic environment. Fig. \ref{fig:BS-WFs}c shows a contour plot of a wave function at the Fermi level (indicated by a red circle on the band structure plot). The wave function is mainly located on the Mo atoms of the structure and its orbital character is $d_{z^2}$-like.

As previously discussed, the initial pool of parent monolayers includes only non-magnetic materials, with the
exception of \ce{VS2} and \ce{CrTe2}. After self-intercalation, a magnetic ground state arises as an emergent property for the
five different, initially non-magnetic, ic-2D materials (see table S1-2): 100\%-ic \ce{TiTe2}, 33\% and 50\%-ic \ce{W2Se4}, 33\%-ic \ce{ZrCl} and 25\%-ic \ce{ZrCl2}. 
With the exception of ic-\ce{TiTe2} and ic-\ce{CrTe2}, where the magnetic moments are uniformly distributed on all the metallic atoms, all the magnetic ic-2D materials have magnetic moments localized on the intercalated atoms. \ce{VS2} preserves its magnetic ground state only at 33\% ic-concentration, where the magnetic moment is shifted from the V atoms of the host bilayer towards the intercalated V atoms. At 67\% concentration the material becomes non-magnetic, i.e. in this case self-intercalation suppresses magnetism.
The energy gained by including spin polarization effects is between 40 and 440 meV per intercalated atom in all the magnetic ic-2D materials, except for ic-\ce{CrTe2}, where the 25\% and 50\%-ic configurations gain 8.0 and 5.1 eV, respectively.
We mention that Zhao \textit{et al.} found the 33\%-ic \ce{TaS2} to be ferromagnetic\cite{zhao2020engineering} while our PBE calculations predict it as non-magnetic. This likely stems from the use of an optimized Hubbard \textit{U}-parameter for Ta atoms in their work.

The significant changes in the electronic properties induced by self-intercalation suggests that the ic-2D materials could complement the original 2D materials in useful ways. Importantly, the pristine and self-intercalated materials share the same chemical elements. In addition, the lattice constants are very similar (see Fig. S4). Thus the pristine 2D material and the ic-2D counterparts are chemically and structurally highly compatible. This could be useful for construction of 2D electronic devices, particular, field effect transistors. For example, one could imagine using a metallic ic-2D material as an electrical contact to its semiconducting pristine counterpart. The high degree of compatibility of the two types of materials are expected to yield high quality interfaces, which is key in order to minimise the contact resistance.   

\paragraph{Catalytic properties}

 \begin{figure}[ht!]
        \centering
        \includegraphics[width=0.9\textwidth]{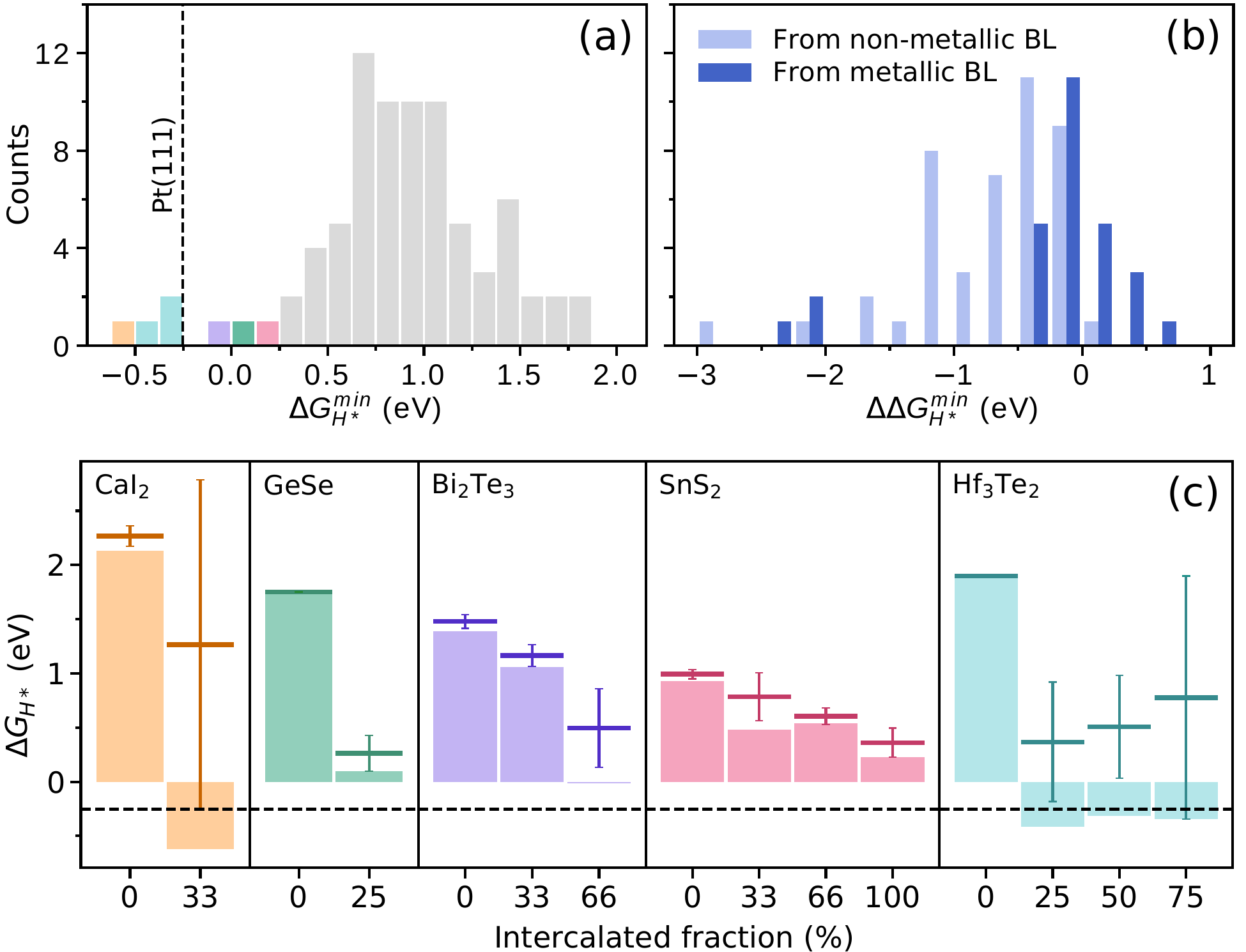}
        \caption{
            a) Distribution of the hydrogen adsorption energies on the 
            lowest-energy site ($\Delta G^{\mathrm{min}}_{H*}$) for the thermodynamically stable ic-2D. 
            The colored bars correspond to the four ic-2D with
            $\Delta G^{\mathrm{min}}_{H*}$ within 0.3 eV from the one calculated for a
            Pt(111) surface (-0.25 eV, dashed line).
            b) Distribution of the $\Delta G^{\mathrm{min}}_{H*}$ of the ic-2D relative to the
            ones calculated for the pristine bilayers.
            light blue and dark blue bars gather the ic-2D obtained from
            non-metallic and metallic bilayers, respectively.
            c) Hydrogen adsorption energies $\Delta G_{H*}$ as a function of the
            ic-concentration. We selected the
            four systems having at least one ic-2D configuration that satisfies 
            the criterion described for panel a).
            The filled bars represent $\Delta G^{\mathrm{min}}_{H*}$, while the darker, horizontal lines
            correspond to the H adsorption energy averaged between all
            the adsorption sites. The error bars show the corresponding standard deviation.
            The horizontal dashed line shows, again, $\Delta G^{\mathrm{min}}_{H*}$ of Pt(111).
            Colors are consistent with panel a).
        }
        \label{fig:adsorption}
    \end{figure}
    
As shown in the previous section, self-intercalation can significantly affect the electronic properties of the host system, e.g. by reducing the band gap of an insulating/semiconducting material, turning it into a metal, or increasing the density of states at the Fermi level of a metal.
Since the reactivity of a material is directly related to its (surface) electronic structure, it is anticipated that the presence of intercalated metal atoms will affect the ability of the material to bind atoms or molecules at the basal plane. Ideally, intercalation could then be used to enhance the catalytic properties of an otherwise chemically inert 2D material. This hypothesis is explored in the following.

To investigate how self-intercalation can affect surface reactivity, we computed the hydrogen adsorption free energy, $\Delta G_{\mathrm{H}^*}$, on the 95 most stable ic-2D materials.
It has been shown by Nørskov
\textit{et al.}\cite{norskov2004origin,montoya2017materials} that $\Delta G_{\mathrm{H}^*}$ correlates well with the catalytic activity of a material towards the hydrogen evolution reaction (HER). More specifically, HER catalysts with higher exchange current densities and lower overpotentials tend to have $\Delta G_{\mathrm{H}^*}$ closer to zero\cite{norskov2004origin}.

For each of the 95 selected ic-2D structures, we determined all the symmetry-inequivalent adsorption sites on the surface and calculated the
corresponding hydrogen adsorption energies. The Methods section provides detailed information on how $\Delta G_{\mathrm{H}^*}$ is calculated.
We expect the site with the lowest 
$\Delta G_{\mathrm{H}^*}$ to dominate the overall reaction dynamics, and we denote the corresponding H adsorption energy as $\Delta G_{\mathrm{H}^*}^{\mathrm{min}}$. 
Fig. \ref{fig:adsorption}a shows the distribution of the calculated $\Delta G_{\mathrm{H}^*}^{\mathrm{min}}$
for the 95 ic-2D structures. The distribution is centered around 1.0 eV, far above the optimal, thermo-neutral adsorption required for HER catalysts.
Nevertheless, seven ic-2D structures are found to be promising candidates for the HER, with $\Delta G_{\mathrm{H}^*}^{\mathrm{min}}$ lying within 0.5 eV of the value obtained for a Pt(111) surface (-250 meV), here used as a model for the ideal HER catalyst. These seven systems are \ce{CaI2} (33\%-ic), \ce{GeSe} (25\%-ic), \ce{Bi2Te3} (66\%-ic), \ce{SnS2} (100\%-ic) and \ce{Hf3Te2} (25, 50 and 75\%-ic).

Fig. \ref{fig:adsorption}b shows the change in $\Delta G^{\mathrm{min}}_{\mathrm{H}^*}$ between the ic-2D structures and their parent pristine bilayers. 
The distribution points towards a general increase in the chemical reactivity, i.e. more negative adsorption energies, 
for ic-2D structures obtained from non-metallic host systems (red bars). This observation correlates well with the observed enhanced metallicity, which is expected to increase the chemical reactivity. For intrinsically metallic host materials (blue bars), self-intercalation seems to leave the reactivity
either unaffected or slightly decreased. One prominent exception is ic-\ce{Hf3Te2} for which  
$\Delta G^{\mathrm{min}}_{\mathrm{H}^*}$ is more than 2 eV lower compared to the pristine bilayer, and in fact constitutes one of the most 
promising HER electro-catalysts found in this work.

The drastic change in $\Delta G^{\mathrm{min}}_{\mathrm{H}^*}$ observed for ic-\ce{Hf3Te2} is ascribed to a particular adsorption configuration in which H is able to 
penetrate the surface and bond to a Hf surface atom from an interstitial site with no energy barrier. 
A similar configuration was found for 33\%-ic \ce{CaI2}, which also showed a comparable change in $\Delta G^{\mathrm{min}}_{\mathrm{H}*}$ relative to the pristine bilayer. 
The two atomic structures are shown in Fig. S5. Interestingly, these special adsorption sites are activated by the self-intercalation. For the pristine bilayers these interstitial adsorption sites are also found upon relaxation, but the H adsorption energy is significantly higher and similar to those at the surface sites.

Fig. \ref{fig:adsorption}c shows the hydrogen adsorption energies as a function of the intercalated fraction 
for \ce{CaI2}, \ce{GeSe}, \ce{Bi2Te3}, \ce{SnS2} and \ce{Hf3Te2}. The colored bars represent $\Delta G^{\mathrm{min}}_{\mathrm{H}*}$ while the horizontal (vertical) lines represent the average (standard deviation) of the distribution of $\Delta G_{\mathrm{H}*}$ over all adsorption sites.  
The variation in $\Delta G_{\mathrm{H}*}$ among the different adsorption sites highlights how the presence of the intercalated atoms affects the 
local environment, resulting in a number of diverse and generally more reactive adsorption sites rather than the few,
chemically equivalent sites present on the pristine host systems.
In Ref. \cite{yang2019ultrahigh}, Yang \textit{et al.} found that self-intercalated \ce{NbS2} (H-phase) presented a significantly enhanced current density in electrocatalytic HER measurements as compared to the non-intercalated \ce{NbS2}. This enhancement was explained by DFT calculations showing hydrogen adsorption energies close to thermoneutrality ($\Delta G^{\mathrm{min}}_{\mathrm{H}*}$) for the Nb-terminated surface of the intercalated system.  We point out that Yang \textit{et al.} report \ce{Nb_{1.35}S2} as the experimentally observed stoichiometry, corresponding to 33\% intercalation. However, their simulated structures seem to correspond to 100\% intercalation. When comparing their reported hydrogen adsorption energy for the S-terminated surface with our 100\%-ic configuration, we obtain a perfect agreement (1.00 eV and 0.997 eV, respectively). 

The theoretical predictions presented by Yang \textit{et al.} are intriguing, as they point towards the Nb-terminated surface being responsible for the observed enhancement in HER reactivity upon intercalation. Consequently, a more complete description of the HER on self-intercalated \ce{NbS2} and other similar ic-2D structures should invoke an analysis of the relative stability of different surface terminations/compositions under the relevant electrochemical conditions.

\section{Conclusions}
Our work shows that self-intercalation opens new possibilities for creating novel types of 2D materials with enhanced metallicity and basal plane activity and potential applications e.g. in (opto)electronics and catalysis.
In accordance with most experiments on self-intercalated 2D materials (ic-2D), the structures investigated in this work possess a period arrangement of the intercalated atoms and corresponding well defined stoichiometries. Thus the considered ic-2D structures represent unique, crystalline, and covalently bonded 2D materials. Consequently, many of their basic properties were calculated and included in the open C2DB database.
Several new stable ic-2D materials were identified by the high-throughput computational workflow. It was shown that self-intercalation generally promotes metallicity and enhances the chemical reactivity of the materials, and a number of ic-2D materials were found to be promising as HER electrocatalysts based on the calculated hydrogen adsorption energy. Future work in this direction should explore the stability of different surface terminations/compositions of the ic-2D materials under HER conditions (e.g. their surface Pourbaix diagrams) as well as the reaction kinetics, possibly with the support of experimental observations. It would also be interesting to examine the assumption of crystalline intercalation structures at finite temperatures and investigate their phase diagram (order-disorder transition), e.g. using Monte Carlo simulations.

\section{Methods}

\paragraph{Density functional theory calculations}
All the DFT calculations are performed with the projector augmented wave (PAW) method\cite{blochl1994projector,kresse1999ultrasoft} as implemented in the GPAW
electronic structure code\cite{enkovaara2010electronic}, using the Perdew-Burke-Ernzerhof (PBE) exchange-correlation functional\cite{perdew1996generalized}. 
Spin polarization is always taken into account and a Fermi smearing of 0.05 eV is consistently applied.
Structure optimizations are carried out in two steps, starting with a coarse 
relaxation of both the unit cell and the atomic positions until the forces are below 0.3 eV \AA$^{-1}$,
followed by a refinement of the atomic positions only until all forces are below 0.03 eV \AA$^{-1}$. 
For relaxations, we use a plane-wave cutoff of 500 eV and a k-point density of 4 \AA$^{-1}$. These are increased to 800 eV and 12 \AA$^{-1}$,
respectively, for electronic ground state calculations.

When relaxing hydrogen adsorption configurations, D3 corrections\cite{grimme2010consistent} are applied in order to take into account the van der Waals interactions.
The D3 corrections are not applied when relaxing the ic-2D structures for the C2DB database for two reasons:
(1) for consistency with the computational parameters used in the C2DB\cite{haastrup2018computational}.
(2) the inter-layer bonding in ic-2D structures is dominated by covalent interactions between the host material layers and the intercalated atoms, rather than dispersion interactions. 
In general, the difference between the thickness of ic-2D structures relaxed with and without D3 corrections is below or slightly above 5\% (and similarly for the in-plane lattice constants), as seen in Fig. S1. Two notable exceptions are 25\% and 75\%-ic \ce{PdS2}, where van der Waals interactions contribute significantly to the binding between the host monolayers and the intercalated atoms. This causes a decrease of 18\% and 16\%, respectively, in the ic-2D thickness. In-depth studies on these systems should thus include D3 corrections in order to correctly describe their atomic structure and related properties.

Vibrational zero-point energies (ZPE) of the Hydrogen adsorbates on the ic-2D surfaces are calculated by displacing the adsorbed H atoms by $\pm 0.01$ \AA ~ along each direction while keeping all other atoms fixed.

\paragraph{Structure generation and workflow}
The ic-2D structures are generated using the Atomic Simulation Environment (ASE)\cite{larsen2017atomic} and its defect builder utility\cite{davidsson2023absorption}, which can identify inequivalent hollow and adsorption sites based on the Voronoi tessellation method.
The computational workflow is built using the ASR-MyQueue\cite{gjerding2021atomic,mortensen2020myqueue} framework.

\paragraph{Hydrogen adsorption energetics}
 Hydrogen adsorption on a surface site, denoted by *, can be described by the chemical reaction
    \begin{equation}
        \ce{H+}(aq) ~ + ~ e^- ~ + ~ * ~ \rightleftharpoons ~ \ce{H}^*
    \label{eq:1}
    \end{equation}
where $\ce{H}^*$ denotes a neutral hydrogen atom adsorbed on the surface site. Overall, the HER process is governed by the electrochemical reduction of solvated protons into gaseous hydrogen:
    \begin{equation}
        \ce{H+}(aq) ~ + ~ e^- ~ \rightleftharpoons ~ 1/2 ~ \ce{H2}(g)
    \label{eq:2}
    \end{equation}
In standard conditions (pH 0, $p(\ce{H2})$ = 1 bar) and at $U=0$ vs. SHE, this process is at equilibrium ($\Delta G^0 = 0$), leading to the equivalence
    \begin{equation}
        \mu^0(\ce{H+}) + \mu^0(e^-) = 1/2 ~ \mu^0(\ce{H2})
    \label{eq:3}
    \end{equation}
Where $\mu^0(i)$ represents the standard chemical potential of species \textit{i} and will be represented herein by its calculated DFT total energy. This allows one to evaluate the Gibbs free energy of reaction \ref{eq:1} (in the same conditions) as
    \begin{equation}
        \Delta G_{\mathrm{H}^*} = E^{DFT}(\ce{H}^*) - E^{DFT}(*)  - 1/2 ~ E^{DFT}(\ce{H2}) + \Delta \mathrm{ZPE} - T \Delta S
    \label{eq:4}
    \end{equation}
Where $E^{DFT}(*)$ and $E^{DFT}(\mathrm{H}^*)$ represent the calculated energy of the pristine surface and the surface with a H atom adsorbed, respectively. $\Delta \mathrm{ZPE}$ is the difference between the zero-point energies of an adsorbed hydrogen atom and hydrogen in the gas phase, while $\Delta S$ is the difference in their standard entropy. We neglect any entropic contribution coming from the solid surfaces and we  assume the entropy of an adsorbed H atom to be small compared to the one in gas phase, hence we approximate $\Delta S \simeq 1/2 ~ S^0(\ce{H2}(g))$. We use
$\mathrm{ZPE}(\ce{H2})$=0.27 eV from Irikura\cite{irikura2007experimental} and $S^0(\ce{H2}(g))$=0.41 eV from Chase \textit{et al.}\cite{chase1998nist}.
The calculated total energy for a \ce{H2} molecule is -6.8 eV.

\section{Data Availability}
The results reported in this article are freely available on the \href{https://cmr.fysik.dtu.dk/c2db/c2db.html}{C2DB} website.

\section{Acknowledgements}
     We acknowledge funding from the European Research Council (ERC) under the European Union’s Horizon 2020 research and innovation program Grant No. 773122 (LIMA) and Grant agreement No. 951786 (NOMAD CoE). K. S. T. is a Villum Investigator supported by VILLUM FONDEN (grant no. 37789).

\section{Supporting Information}
The provided supplementary material contains information about the effect of (a) self-intercalation on the lattice structure, (b) atomic composition on the formation energy. (c) D3 corrections on the ic-2D structure. Atomic structure referenced in the article text are also provided. Finally, we include a table summarizing the main thermodynamic and electronic properties of all the 95 stable ic-2D.

\bibliographystyle{unsrt}
\bibliography{references}

\begin{thebibliography}{10}

\bibitem{das2021transistors}
Saptarshi Das, Amritanand Sebastian, Eric Pop, Connor~J McClellan, Aaron~D Franklin, Tibor Grasser, Theresia Knobloch, Yury Illarionov, Ashish~V Penumatcha, Joerg Appenzeller, et~al.
\newblock Transistors based on two-dimensional materials for future integrated circuits.
\newblock {\em Nature Electronics}, 4(11):786--799, 2021.

\bibitem{cao20212d}
Guiming Cao, Peng Meng, Jiangang Chen, Haishi Liu, Renji Bian, Chao Zhu, Fucai Liu, and Zheng Liu.
\newblock 2d material based synaptic devices for neuromorphic computing.
\newblock {\em Advanced Functional Materials}, 31(4):2005443, 2021.

\bibitem{ahn20202d}
Ethan~C Ahn.
\newblock 2d materials for spintronic devices.
\newblock {\em npj 2D Materials and Applications}, 4(1):17, 2020.

\bibitem{wang2023towards}
Chuanshou Wang, Lu~You, David Cobden, and Junling Wang.
\newblock Towards two-dimensional van der waals ferroelectrics.
\newblock {\em Nature Materials}, pages 1--11, 2023.

\bibitem{lee2019chemical}
Chung~Won Lee, Jun~Min Suh, and Ho~Won Jang.
\newblock Chemical sensors based on two-dimensional (2d) materials for selective detection of ions and molecules in liquid.
\newblock {\em Frontiers in Chemistry}, 7:708, 2019.

\bibitem{anichini2018chemical}
Cosimo Anichini, W{\l}odzimierz Czepa, Dawid Pakulski, Alessandro Aliprandi, Artur Ciesielski, and Paolo Samor{\`\i}.
\newblock Chemical sensing with 2d materials.
\newblock {\em Chemical Society Reviews}, 47(13):4860--4908, 2018.

\bibitem{cummins2016efficient}
Dustin~R Cummins, Ulises Martinez, Andriy Sherehiy, Rajesh Kappera, Alejandro Martinez-Garcia, Roland~K Schulze, Jacek Jasinski, Jing Zhang, Ram~K Gupta, Jun Lou, et~al.
\newblock Efficient hydrogen evolution in transition metal dichalcogenides via a simple one-step hydrazine reaction.
\newblock {\em Nature communications}, 7(1):11857, 2016.

\bibitem{pandey2017two}
Mohnish Pandey and Kristian~S Thygesen.
\newblock Two-dimensional mxenes as catalysts for electrochemical hydrogen evolution: A computational screening study.
\newblock {\em The Journal of Physical Chemistry C}, 121(25):13593--13598, 2017.

\bibitem{pandey2015two}
Mohnish Pandey, Aleksandra Vojvodic, Kristian~S Thygesen, and Karsten~W Jacobsen.
\newblock Two-dimensional metal dichalcogenides and oxides for hydrogen evolution: a computational screening approach.
\newblock {\em The journal of physical chemistry letters}, 6(9):1577--1585, 2015.

\bibitem{li20192d}
Zhe Li and Yue Wu.
\newblock 2d early transition metal carbides (mxenes) for catalysis.
\newblock {\em Small}, 15(29):1804736, 2019.

\bibitem{zhang2018single}
Jinqiang Zhang, Yufei Zhao, Xin Guo, Chen Chen, Chung-Li Dong, Ru-Shi Liu, Chih-Pin Han, Yadong Li, Yury Gogotsi, and Guoxiu Wang.
\newblock Single platinum atoms immobilized on an mxene as an efficient catalyst for the hydrogen evolution reaction.
\newblock {\em Nature Catalysis}, 1(12):985--992, 2018.

\bibitem{mondal20222d}
Aniruddha Mondal and Alberto Vomiero.
\newblock 2d transition metal dichalcogenides-based electrocatalysts for hydrogen evolution reaction.
\newblock {\em Advanced Functional Materials}, 32(52):2208994, 2022.

\bibitem{han2018activation}
Jae~Hyo Han, Hong~Ki Kim, Bongkwan Baek, Jeonghee Han, Hyun~S Ahn, Mu-Hyun Baik, and Jinwoo Cheon.
\newblock Activation of the basal plane in two dimensional transition metal chalcogenide nanostructures.
\newblock {\em Journal of the American Chemical Society}, 140(42):13663--13671, 2018.

\bibitem{yang2019ultrahigh}
Jieun Yang, Abdul~Rahman Mohmad, Yan Wang, Raymond Fullon, Xiuju Song, Fang Zhao, Ibrahim Bozkurt, Mathias Augustin, Elton~JG Santos, Hyeon~Suk Shin, et~al.
\newblock Ultrahigh-current-density niobium disulfide catalysts for hydrogen evolution.
\newblock {\em Nature Materials}, 18(12):1309--1314, 2019.

\bibitem{jariwala2014emerging}
Deep Jariwala, Vinod~K Sangwan, Lincoln~J Lauhon, Tobin~J Marks, and Mark~C Hersam.
\newblock Emerging device applications for semiconducting two-dimensional transition metal dichalcogenides.
\newblock {\em ACS Nano}, 8(2):1102--1120, 2014.

\bibitem{briggs2019roadmap}
Natalie Briggs, Shruti Subramanian, Zhong Lin, Xufan Li, Xiaotian Zhang, Kehao Zhang, Kai Xiao, David Geohegan, Robert Wallace, Long-Qing Chen, et~al.
\newblock A roadmap for electronic grade 2d materials.
\newblock {\em 2D Materials}, 6(2):022001, 2019.

\bibitem{peimyoo2021electrical}
Namphung Peimyoo, Thorsten Deilmann, Freddie Withers, Janire Escolar, Darren Nutting, Takashi Taniguchi, Kenji Watanabe, Alireza Taghizadeh, Monica~Felicia Craciun, Kristian~Sommer Thygesen, et~al.
\newblock Electrical tuning of optically active interlayer excitons in bilayer mos2.
\newblock {\em Nature Nanotechnology}, 16(8):888--893, 2021.

\bibitem{tran2019evidence}
Kha Tran, Galan Moody, Fengcheng Wu, Xiaobo Lu, Junho Choi, Kyounghwan Kim, Amritesh Rai, Daniel~A Sanchez, Jiamin Quan, Akshay Singh, et~al.
\newblock Evidence for moir{\'e} excitons in van der waals heterostructures.
\newblock {\em Nature}, 567(7746):71--75, 2019.

\bibitem{winther2017band}
Kirsten~T Winther and Kristian~S Thygesen.
\newblock Band structure engineering in van der waals heterostructures via dielectric screening: the g$\delta$w method.
\newblock {\em 2D Materials}, 4(2):025059, 2017.

\bibitem{rajapakse2021intercalation}
Manthila Rajapakse, Bhupendra Karki, Usman~O Abu, Sahar Pishgar, Md~Rajib~Khan Musa, SM~Shah Riyadh, Ming Yu, Gamini Sumanasekera, and Jacek~B Jasinski.
\newblock Intercalation as a versatile tool for fabrication, property tuning, and phase transitions in 2d materials.
\newblock {\em npj 2D Materials and Applications}, 5(1):30, 2021.

\bibitem{wu2023electrostatic}
Yecun Wu, Danfeng Li, Chun-Lan Wu, Harold~Y Hwang, and Yi~Cui.
\newblock Electrostatic gating and intercalation in 2d materials.
\newblock {\em Nature Reviews Materials}, 8(1):41--53, 2023.

\bibitem{zhang2023mechanism}
Peikun Zhang, Minmin Xue, Changfeng Chen, Wanlin Guo, and Zhuhua Zhang.
\newblock Mechanism regulating self-intercalation in layered materials.
\newblock {\em Nano Letters}, 23(8):3623--3629, 2023.

\bibitem{zhao2020engineering}
Xiaoxu Zhao, Peng Song, Chengcai Wang, Anders~C Riis-Jensen, Wei Fu, Ya~Deng, Dongyang Wan, Lixing Kang, Shoucong Ning, Jiadong Dan, et~al.
\newblock Engineering covalently bonded 2d layered materials by self-intercalation.
\newblock {\em Nature}, 581(7807):171--177, 2020.

\bibitem{cui2020controlled}
Fangfang Cui, Xiaoxu Zhao, Junjie Xu, Bin Tang, Qiuyu Shang, Jianping Shi, Yahuan Huan, Jianhui Liao, Qing Chen, Yanglong Hou, et~al.
\newblock Controlled growth and thickness-dependent conduction-type transition of 2d ferrimagnetic cr2s3 semiconductors.
\newblock {\em Advanced Materials}, 32(4):1905896, 2020.

\bibitem{li2019molecular}
Hongxi Li, Linjing Wang, Junshu Chen, Tao Yu, Liang Zhou, Yang Qiu, Hongtao He, Fei Ye, Iam~Keong Sou, and Gan Wang.
\newblock Molecular beam epitaxy grown cr2te3 thin films with tunable curie temperatures for spintronic devices.
\newblock {\em ACS Applied Nano Materials}, 2(11):6809--6817, 2019.

\bibitem{wen2020tunable}
Yao Wen, Zhehong Liu, Yu~Zhang, Congxin Xia, Baoxing Zhai, Xinhui Zhang, Guihao Zhai, Chao Shen, Peng He, Ruiqing Cheng, et~al.
\newblock Tunable room-temperature ferromagnetism in two-dimensional cr2te3.
\newblock {\em Nano Letters}, 20(5):3130--3139, 2020.

\bibitem{lasek2022van}
Kinga Lasek, Paula~M Coelho, Pierluigi Gargiani, Manuel Valvidares, Katayoon Mohseni, Holger~L Meyerheim, Ilya Kostanovskiy, Krzysztof Zberecki, and Matthias Batzill.
\newblock Van der waals epitaxy growth of 2d ferromagnetic cr(1+ $\delta$)te2 nanolayers with concentration-tunable magnetic anisotropy.
\newblock {\em Applied Physics Reviews}, 9(1):011409, 2022.

\bibitem{lasek2020molecular}
Kinga Lasek, Paula~Mariel Coelho, Krzysztof Zberecki, Yan Xin, Sadhu~K Kolekar, Jingfeng Li, and Matthias Batzill.
\newblock Molecular beam epitaxy of transition metal (ti-, v-, and cr-) tellurides: From monolayer ditellurides to multilayer self-intercalation compounds.
\newblock {\em ACS nano}, 14(7):8473--8484, 2020.

\bibitem{li2021magnetic}
Qiu-Qiu Li, Si~Li, Dan Wu, Zhong-Ke Ding, Xuan-Hao Cao, Lin Huang, Hui Pan, Bo~Li, Ke-Qiu Chen, and Xi-Dong Duan.
\newblock Magnetic properties manipulation of crte2 bilayer through strain and self-intercalation.
\newblock {\em Applied Physics Letters}, 119(16):162402, 2021.

\bibitem{li2022tailoring}
Rui Li, Jiawei Jiang, Haili Bai, and Wenbo Mi.
\newblock Tailoring interlayer magnetic coupling to modify the magnetic properties of fecl 2 bilayers by self-intercalation.
\newblock {\em Journal of Materials Chemistry C}, 10(40):14955--14962, 2022.

\bibitem{guo2020enhanced}
Yu~Guo, Nanshu Liu, Yanyan Zhao, Xue Jiang, Si~Zhou, and Jijun Zhao.
\newblock Enhanced ferromagnetism of cri3 bilayer by self-intercalation.
\newblock {\em Chinese Physics Letters}, 37(10):107506, 2020.

\bibitem{haastrup2018computational}
Sten Haastrup, Mikkel Strange, Mohnish Pandey, Thorsten Deilmann, Per~S Schmidt, Nicki~F Hinsche, Morten~N Gjerding, Daniele Torelli, Peter~M Larsen, Anders~C Riis-Jensen, et~al.
\newblock The computational 2d materials database: high-throughput modeling and discovery of atomically thin crystals.
\newblock {\em 2D Materials}, 5(4):042002, 2018.

\bibitem{gjerding2021recent}
Morten~Niklas Gjerding, Alireza Taghizadeh, Asbj{\o}rn Rasmussen, Sajid Ali, Fabian Bertoldo, Thorsten Deilmann, Nikolaj~R{\o}rb{\ae}k Kn{\o}sgaard, Mads Kruse, Ask~Hjorth Larsen, Simone Manti, et~al.
\newblock Recent progress of the computational 2d materials database (c2db).
\newblock {\em 2D Materials}, 8(4):044002, 2021.

\bibitem{bergerhoff1987crystallographic}
G~Bergerhoff, ID~Brown, F~Allen, et~al.
\newblock Crystallographic databases.
\newblock {\em International Union of Crystallography, Chester}, 360:77--95, 1987.

\bibitem{gravzulis2012crystallography}
Saulius Gra{\v{z}}ulis, Adriana Da{\v{s}}kevi{\v{c}}, Andrius Merkys, Daniel Chateigner, Luca Lutterotti, Miguel Quiros, Nadezhda~R Serebryanaya, Peter Moeck, Robert~T Downs, and Armel Le~Bail.
\newblock Crystallography open database (cod): an open-access collection of crystal structures and platform for world-wide collaboration.
\newblock {\em Nucleic acids research}, 40(D1):D420--D427, 2012.

\bibitem{taylor1964abundance}
S~Re Taylor.
\newblock Abundance of chemical elements in the continental crust: a new table.
\newblock {\em Geochimica et cosmochimica acta}, 28(8):1273--1285, 1964.

\bibitem{pakdel2023emergent}
Sahar Pakdel, Asbj{\o}rn Rasmussen, Alireza Taghizadeh, Mads Kruse, Thomas Olsen, and Kristian~S Thygesen.
\newblock Emergent properties of van der waals bilayers revealed by computational stacking.
\newblock {\em arXiv preprint arXiv:2304.01148}, 2023.

\bibitem{saal2013materials}
James~E Saal, Scott Kirklin, Muratahan Aykol, Bryce Meredig, and Christopher Wolverton.
\newblock Materials design and discovery with high-throughput density functional theory: the open quantum materials database (oqmd).
\newblock {\em Jom}, 65:1501--1509, 2013.

\bibitem{gossenberger2020sulfate}
Florian Gossenberger, Fernanda Juarez, and Axel Gro{\ss}.
\newblock Sulfate, bisulfate, and hydrogen co-adsorption on pt (111) and au (111) in an electrochemical environment.
\newblock {\em Frontiers in Chemistry}, 8:634, 2020.

\bibitem{yang2019vanadium}
Mingyang Yang, Lujie Cao, Zhenyu Wang, Yuanju Qu, Chaoqun Shang, Hanyu Guo, Wei Xiong, Junjun Zhang, Run Shi, Jianli Zou, et~al.
\newblock Vanadium self-intercalated c/v1. 11s2 nanosheets with abundant active sites for enhanced electro-catalytic hydrogen evolution.
\newblock {\em Electrochimica Acta}, 300:208--216, 2019.

\bibitem{zhang2021deciphering}
Chao~Yue Zhang, Guo~Wen Sun, Qian~Yu Liu, Jiang~Long Pan, Yan~Chun Wang, Hao Zhao, Geng~Zhi Sun, Xiu~Ping Gao, Xiao~Jun Pan, Jin~Yuan Zhou, et~al.
\newblock Deciphering the catalysis essence of vanadium self-intercalated two-dimensional vanadium sulfides (v5s8) on lithium polysulfide towards high-rate and ultra-stable li-s batteries.
\newblock {\em Energy Storage Materials}, 43:471--481, 2021.

\bibitem{coelho2018post}
Paula~Mariel Coelho, Hannu-Pekka Komsa, Horacio Coy~Diaz, Yujing Ma, Arkady~V Krasheninnikov, and Matthias Batzill.
\newblock Post-synthesis modifications of two-dimensional mose2 or mote2 by incorporation of excess metal atoms into the crystal structure.
\newblock {\em ACS nano}, 12(4):3975--3984, 2018.

\bibitem{norskov2004origin}
Jens~Kehlet N{\o}rskov, Jan Rossmeisl, Ashildur Logadottir, LRKJ Lindqvist, John~R Kitchin, Thomas Bligaard, and Hannes Jonsson.
\newblock Origin of the overpotential for oxygen reduction at a fuel-cell cathode.
\newblock {\em The Journal of Physical Chemistry B}, 108(46):17886--17892, 2004.

\bibitem{montoya2017materials}
Joseph~H Montoya, Linsey~C Seitz, Pongkarn Chakthranont, Aleksandra Vojvodic, Thomas~F Jaramillo, and Jens~K N{\o}rskov.
\newblock Materials for solar fuels and chemicals.
\newblock {\em Nature materials}, 16(1):70--81, 2017.

\bibitem{blochl1994projector}
Peter~E Bl{\"o}chl.
\newblock Projector augmented-wave method.
\newblock {\em Physical review B}, 50(24):17953, 1994.

\bibitem{kresse1999ultrasoft}
Georg Kresse and Daniel Joubert.
\newblock From ultrasoft pseudopotentials to the projector augmented-wave method.
\newblock {\em Physical review b}, 59(3):1758, 1999.

\bibitem{enkovaara2010electronic}
Jussi Enkovaara, Carsten Rostgaard, J~J{\o}rgen Mortensen, Jingzhe Chen, M~Du{\l}ak, Lara Ferrighi, Jeppe Gavnholt, Christian Glinsvad, V~Haikola, HA~Hansen, et~al.
\newblock Electronic structure calculations with gpaw: a real-space implementation of the projector augmented-wave method.
\newblock {\em Journal of physics: Condensed matter}, 22(25):253202, 2010.

\bibitem{perdew1996generalized}
John~P Perdew, Kieron Burke, and Matthias Ernzerhof.
\newblock Generalized gradient approximation made simple.
\newblock {\em Physical review letters}, 77(18):3865, 1996.

\bibitem{grimme2010consistent}
Stefan Grimme, Jens Antony, Stephan Ehrlich, and Helge Krieg.
\newblock A consistent and accurate ab initio parametrization of density functional dispersion correction (dft-d) for the 94 elements h-pu.
\newblock {\em The Journal of chemical physics}, 132(15), 2010.

\bibitem{larsen2017atomic}
Ask~Hjorth Larsen, Jens~J{\o}rgen Mortensen, Jakob Blomqvist, Ivano~E Castelli, Rune Christensen, Marcin Du{\l}ak, Jesper Friis, Michael~N Groves, Bj{\o}rk Hammer, Cory Hargus, et~al.
\newblock The atomic simulation environment—a python library for working with atoms.
\newblock {\em Journal of Physics: Condensed Matter}, 29(27):273002, 2017.

\bibitem{davidsson2023absorption}
Joel Davidsson, Fabian Bertoldo, Kristian~S Thygesen, and Rickard Armiento.
\newblock Absorption versus adsorption: high-throughput computation of impurities in 2d materials.
\newblock {\em npj 2D Materials and Applications}, 7(1):1--7, 2023.

\bibitem{gjerding2021atomic}
Morten Gjerding, Thorbj{\o}rn Skovhus, Asbj{\o}rn Rasmussen, Fabian Bertoldo, Ask~Hjorth Larsen, Jens~J{\o}rgen Mortensen, and Kristian~Sommer Thygesen.
\newblock Atomic simulation recipes: A python framework and library for automated workflows.
\newblock {\em Computational Materials Science}, 199:110731, 2021.

\bibitem{mortensen2020myqueue}
Jens~J{\o}rgen Mortensen, Morten Gjerding, and Kristian~Sommer Thygesen.
\newblock Myqueue: Task and workflow scheduling system.
\newblock {\em Journal of Open Source Software}, 5(45):1844, 2020.

\bibitem{irikura2007experimental}
Karl~K Irikura.
\newblock Experimental vibrational zero-point energies: Diatomic molecules.
\newblock {\em Journal of physical and chemical reference data}, 36(2):389--397, 2007.

\bibitem{chase1998nist}
Malcolm~W Chase and National Information Standards~Organization (US).
\newblock {\em NIST-JANAF thermochemical tables}, volume~9.
\newblock American Chemical Society Washington, DC, 1998.

\end{thebibliography}
\end{document}


\title{Functionalised 2D materials: Enhancing metallicity and chemical reactivity \emph{via} self-intercalation\\
\vspace{0.5cm} {\LARGE Supporting information}}
\author[1,*]{Stefano Americo}
\author[1]{Sahar Pakdel}
\author[1]{Kristian Sommer Thygesen}
\affil[1]{Computational Atomic-scale Materials Design (CAMD), Department of Physics, Technical University of Denmark, 2800 Kgs. Lyngby Denmark}
\affil[*]{Email: steame@dtu.dk}

\date{}
\maketitle

\pagebreak

\section*{Effect of van der Waals corrections on the ic-2D structure}
We determine the effect of the D3 correction scheme on the thickness and in-plane lattice constant of the stable ic-2D, by comparing the relaxed structure of all the 95 stable ic-2D obtained with and without corrections.
The thickness is defined as the difference between the largest and the smallest atomic coordinate along the z-axis.
In general, we observe a good agreement between structures optimized with and without D3 corrections. As for the in-plane lattice constant, only 33\%-ic \ce{VS2} shows a discrepancy larger than 5\% (5.04\%). 
Concerning the thickness, the relative difference is below 5\% in almost all cases, with the notable exceptions of \ce{PdSe2} (50\%-ic) and \ce{PdS2} (25\%, 50\% and 75\%-ic).
In particular, 25\% and 75\% intercalated \ce{PdS2} show significant relative differences of 18\% and 16\%, respectively. In these systems, dispersive interactions play a major role in binding the host system monolayers to the intercalated atoms and should be taken into account for an accurate description of their atomic structure.

\begin{figure}[h!]
        \includegraphics[width=\textwidth]{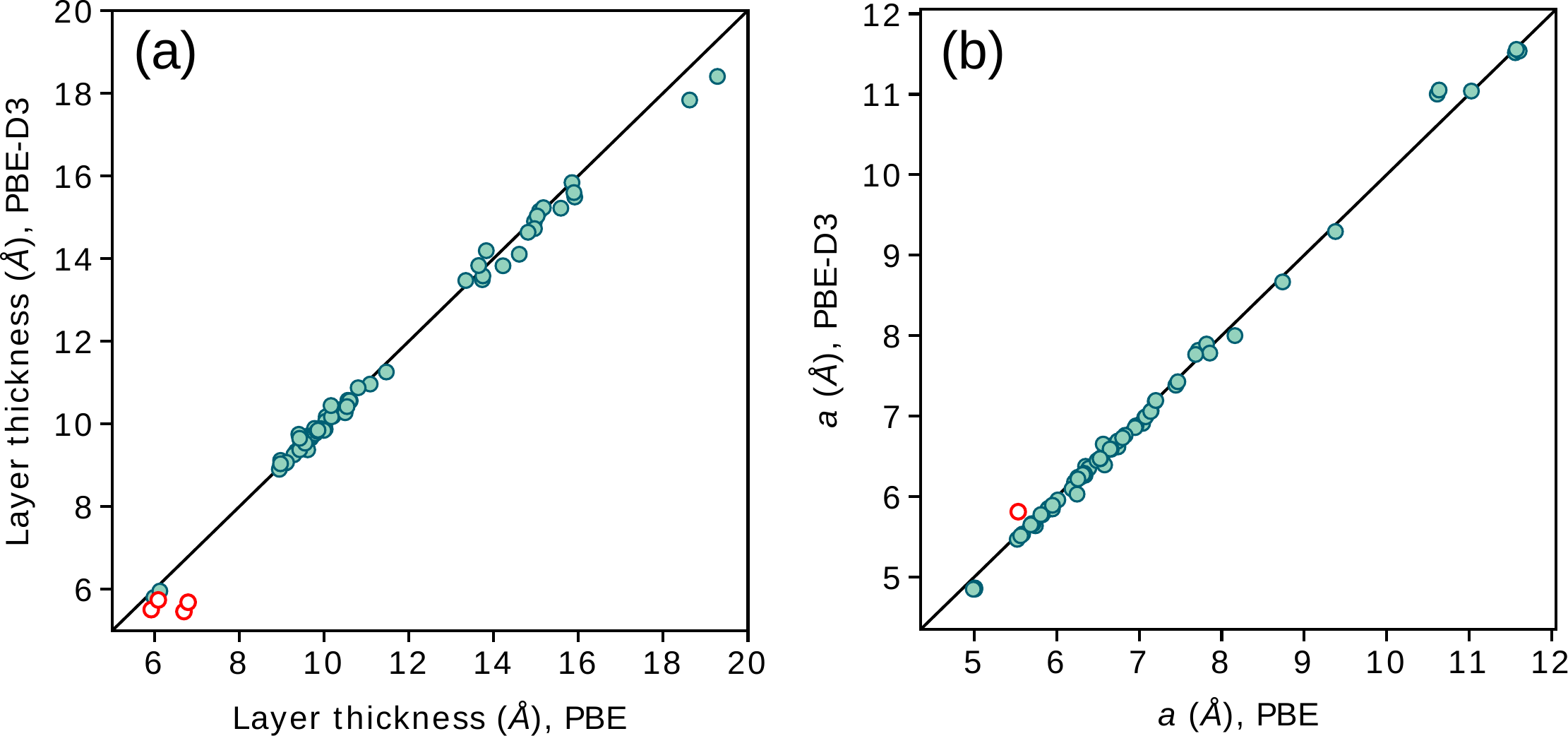}
        \centering
        \caption{
            (a) Comparison between the thickness of ic-2D relaxed with and without D3 corrections.
            ic-2D where the relative difference between such values exceeds 5\% are highlighted with a red circle.
            (b) Comparison between the in-plane lattice constant \textit{a} of ic-2D relaxed with and without D3 corrections. Again, we highlight in red the only case where the relative difference exceeds 5\%.
        }
\end{figure}

\pagebreak

\section*{Structural archetypes of the stable self-intercalated bilayers}
Here we show the six different structural archetypes identified among the 95 stable ic-2D. Note that the archetype is defined by the structure and stoichiometry of the host system in monolayer form and not by the position of the intercalated atoms or the stacking configuration of the host bilayer.

\begin{figure}[h!]
        \vspace{10pt}
        \includegraphics[width=\textwidth]{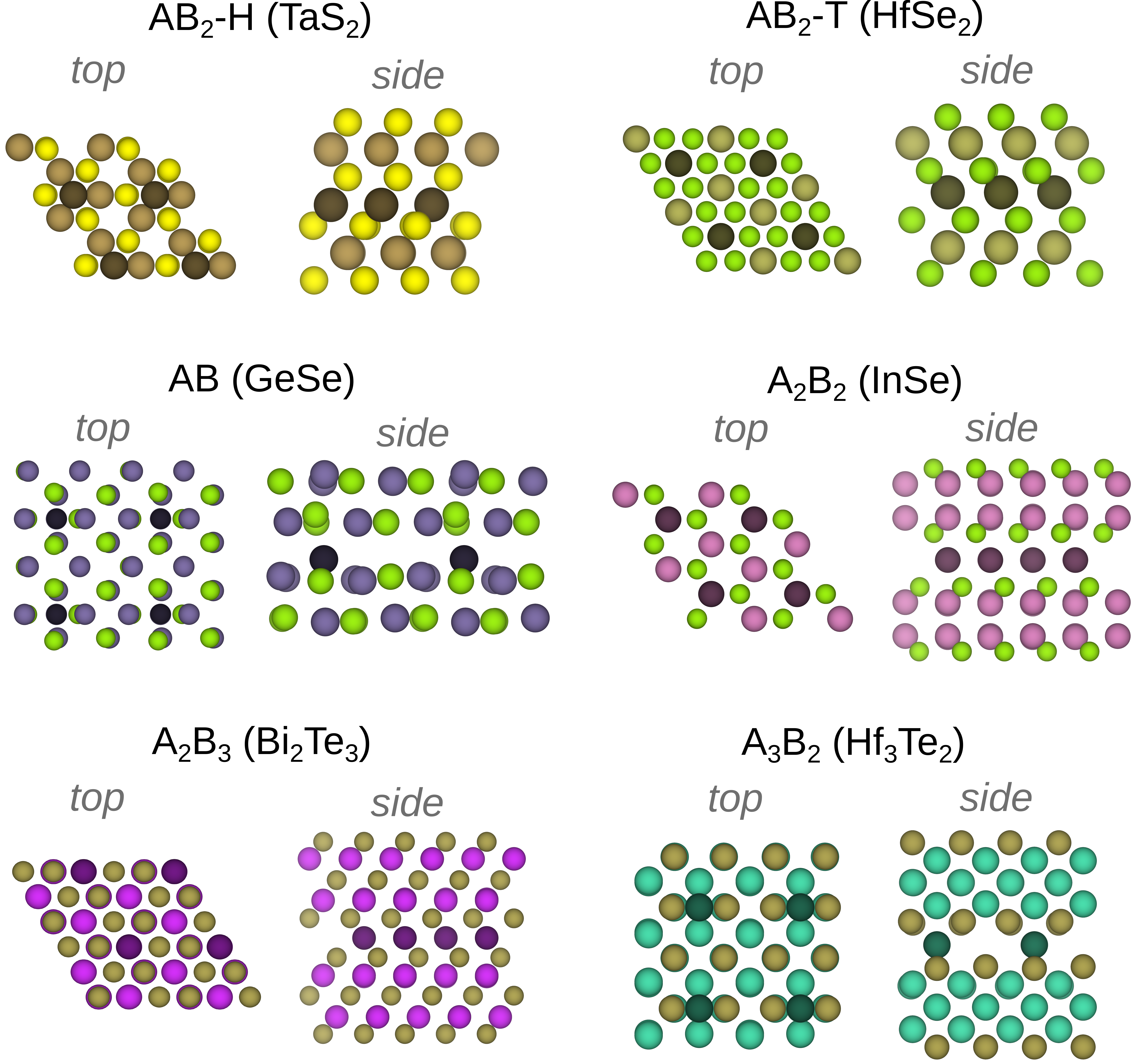}
        \centering
        \caption{
            Top and side views of the six structural archetypes identified among the 95 stable ic-2D. In the top views, the upper layer has been removed in order to highlight the positions of the intercalated atoms. On top of each archetype is reported a label and, in parenthesis, the representative host system of choice.
            All the selected structures are shown in a 2x2 supercell, at either 25\% or 33\% intercalation. Intercalated atoms are shown in darker colors.
        }
\end{figure}

\pagebreak

\section*{Correlation between atomic radii and formation energies}
We investigated whether any descriptor based on the host material composition can be used in order to assess the stability of its intercalated phases. Here, materials with large metallic atoms relative to the non-metallic ones are found to generally produce more negative formation energies.

\begin{figure}[h!]
        \includegraphics[width=0.7\textwidth]{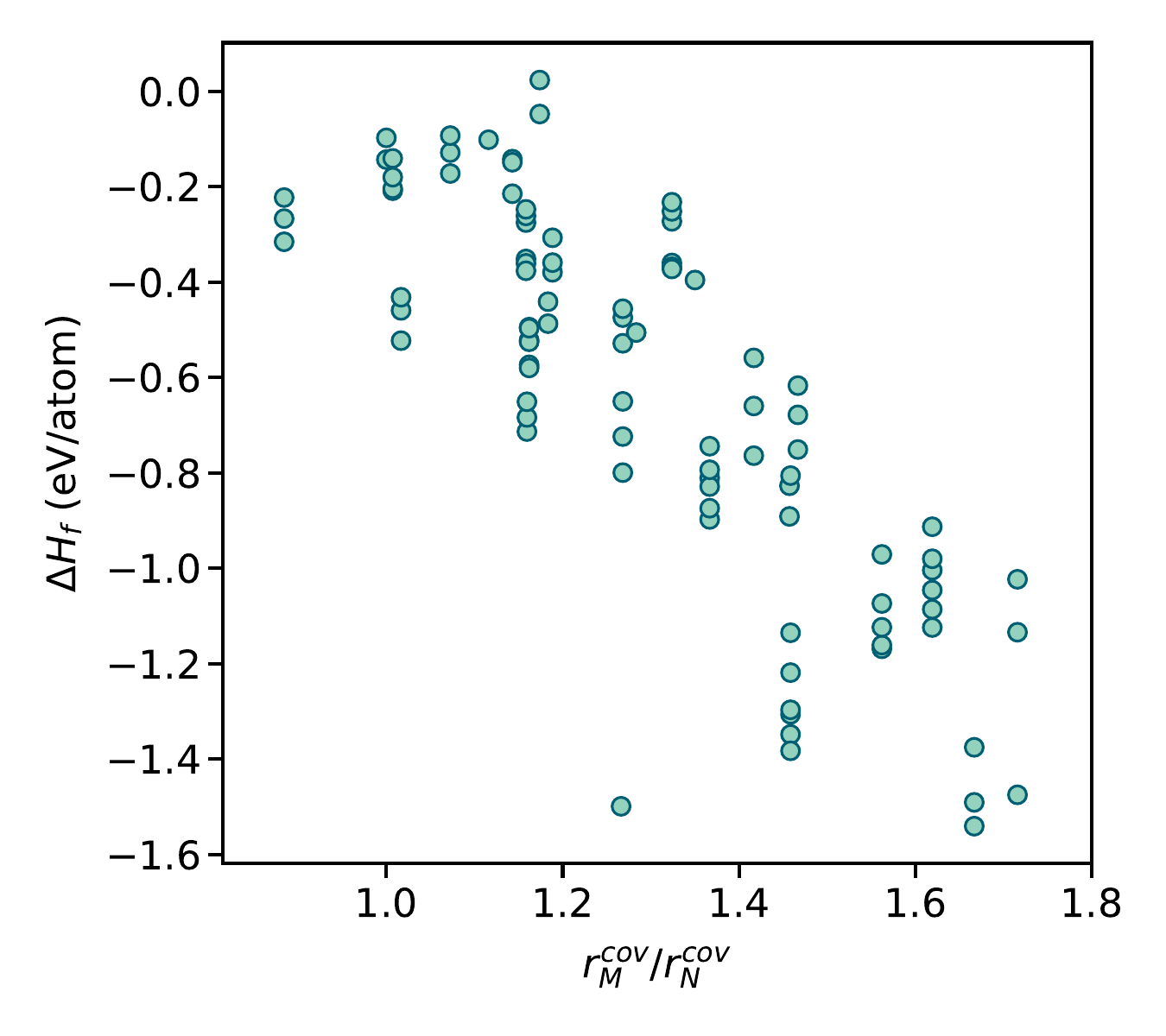}
        \centering
        \caption{
            Formation energy of the stable ic-2D as a function of the ratio between the covalent radii
            of the metallic element ($r_{M}^{cov}$) and the non-metallic element ($r_{N}^{cov}$).
        }
\end{figure}

\pagebreak

\section*{Effect of self-intercalation on the lattice constants}
We compare the first component of the lattice constants between the intercalated phases and their pristine host structure. Self-intercalation does not change the unit cell of the material significantly, with the largest percentage variations being around 4.5\% and the overall distribution centered around 0.

\begin{figure}[h!]
        \includegraphics[width=0.8\textwidth]{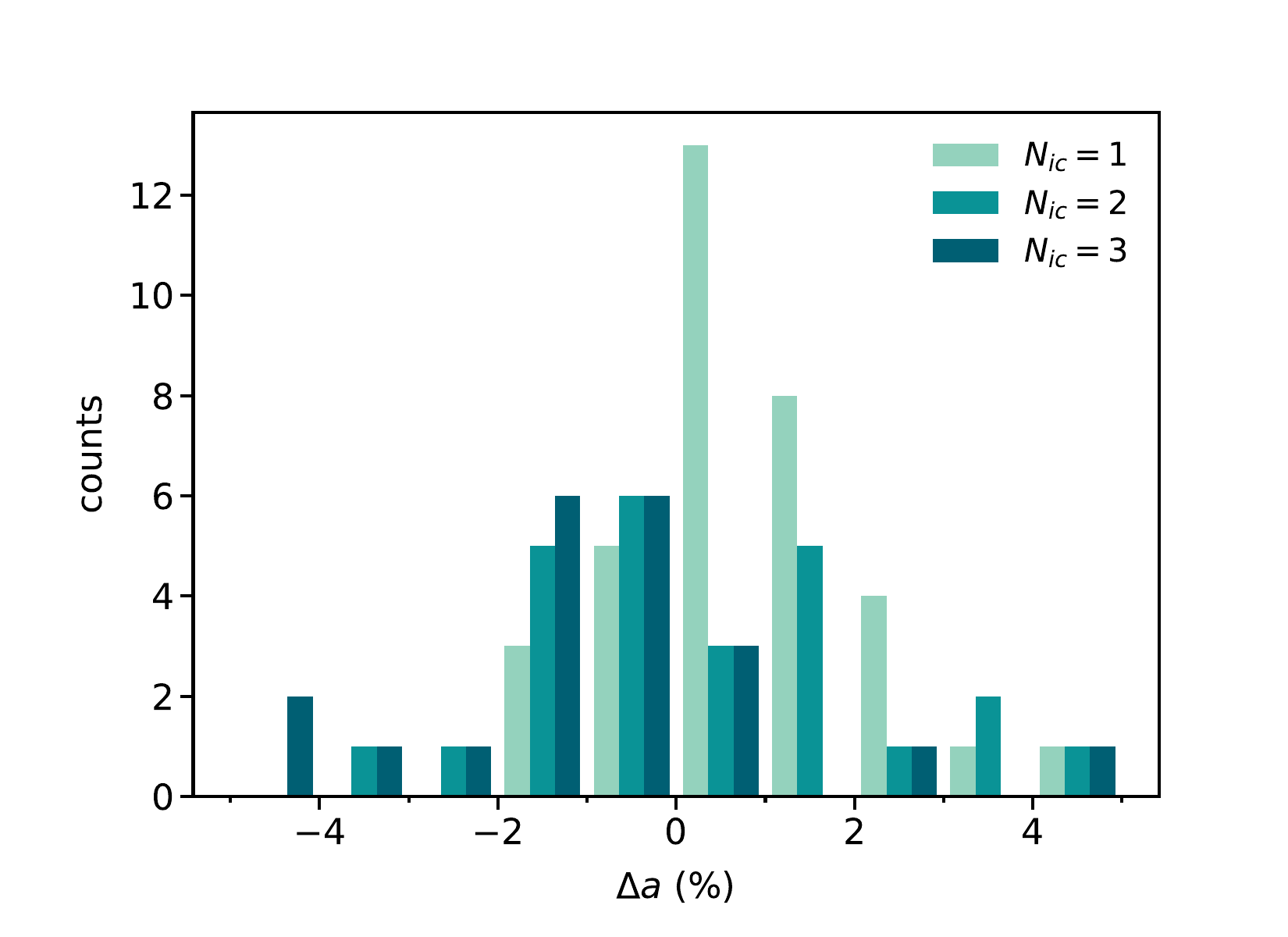}
        \centering
        \caption{
            in-plane lattice constant variation ($\Delta a$) between the ic-2D and their corresponding pristine host materials, expressed as a percentage.
            The data set is subdivided according to the number of intercalated atoms in the ic-2D ($N_{ic}$)
        }
\end{figure}

\pagebreak

\section*{Highly stable H adsorption configurations}
Here we show the structures of two hydrogen adsorption configurations that are significantly stabilized by self-intercalation.

\begin{figure}[h!]
        \includegraphics[width=0.7\textwidth]{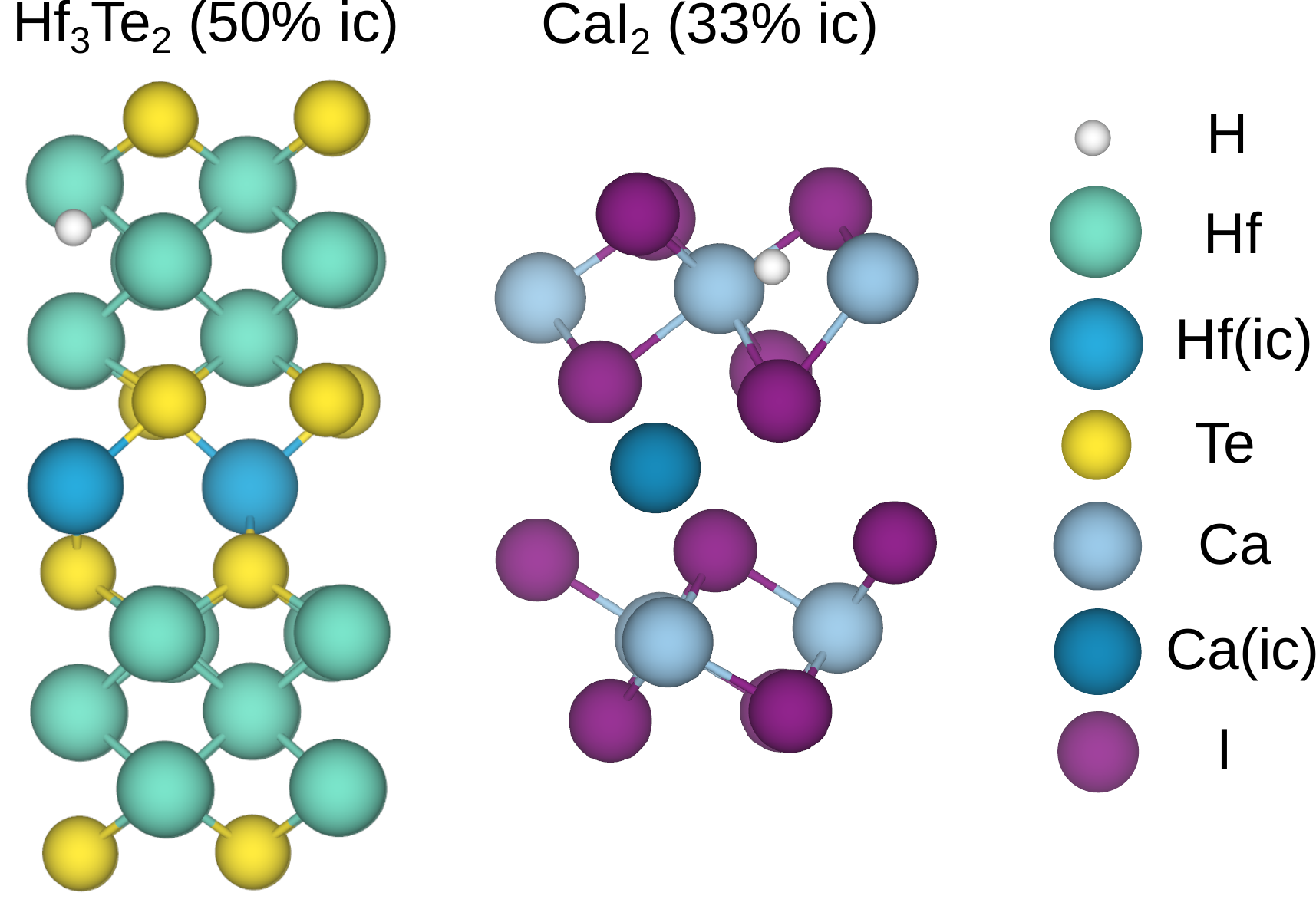}
        \centering
        \caption{
            Hydrogen adsorption configurations providing the lowest $\Delta G^{\mathrm{min}}_{\mathrm{H}*}$ of 50\%-ic \ce{Hf3Te2} and 33\%-ic \ce{CaI2}.
        }
\end{figure}

\pagebreak

\section*{Complete list of stable ic-2D}
\begin{table}[h!]
    \centering
    \resizebox{0.67\textwidth}{!}{\begin{tabular}{lccccc}
\hline\hline
                 Parent ML  &  ic-fraction  &  $\Delta H$ (eV/atom)  &  $\Delta H_{\mathrm{hull}}$ (meV/atom)  &  $E_{\mathrm{gap}}$ (eV) & $\mu_{\mathrm{mag}}$  \\
\hline
\rowcolor{Gray}
  \ce{Bi2Te3}  &    0.33     &     -0.18     &            45.2            &              0               &              0              \\
\rowcolor{Gray}
               &    0.67     &     -0.14     &            76.1            &              0               &              0              \\
\rowcolor{Gray}
               &      1      &     -0.11     &           102.8            &              0               &              0              \\
   \ce{CaI2}   &    0.33     &     -1.5      &            78.9            &             0.01             &              0              \\
\rowcolor{Gray}
   \ce{CrTe2}   &     0.25     &     -0.23     &            22.8             &               0               &            23.70            \\
\rowcolor{Gray}
               &      0.5      &     -0.23     &            39.7            &               0               &            28.74            \\
   \ce{GaSe}   &    0.33     &     -0.53     &             51             &              0               &              0              \\
               &    0.67     &     -0.47     &            81.9            &              0               &              0              \\
               &      1      &     -0.45     &            87.2            &              0               &              0              \\
\rowcolor{Gray}
   \ce{GaS}($\alpha$)    &    0.33     &     -0.58     &            56.4            &              0               &              0              \\
\rowcolor{Gray}
                         &    0.67     &     -0.53     &            84.7            &              0               &              0              \\
\rowcolor{Gray}
                         &      1      &     -0.51     &            84.6            &              0               &              0              \\
   \ce{GaS}($\beta$)    &    0.33     &     -0.58     &            62.5            &              0               &              0              \\
                        &    0.67     &     -0.52     &            89.2            &              0               &              0              \\
                        &      1      &     -0.5      &            88.5            &              0               &              0              \\
\rowcolor{Gray}
   \ce{GaTe}   &    0.33     &     -0.32     &            40.3            &              0               &              0              \\
\rowcolor{Gray}
               &    0.67     &     -0.27     &             76             &              0               &              0              \\
\rowcolor{Gray}
               &      1      &     -0.23     &           105.2            &              0               &              0              \\
   \ce{GeS2}   &    0.75     &     -0.2      &           120.2            &             1.05             &              0              \\
\rowcolor{Gray}
   \ce{GeSe}   &    0.25     &     -0.15     &            68.6            &              0               &              0              \\
\rowcolor{Gray}
               &     0.5     &     -0.1      &           118.5            &              0               &              0              \\
  \ce{Hf3Te2}  &    0.25     &     -0.54     &            54.5            &              0               &              0              \\
               &     0.5     &     -0.48     &            98.1            &              0               &              0              \\
\rowcolor{Gray}
 \ce{HfS2}(T)  &    0.33     &     -1.57     &           -39.2            &              0               &              0              \\
\rowcolor{Gray}
               &    0.67     &     -1.52     &           -34.2            &              0               &              0              \\
\rowcolor{Gray}
               &      1      &     -1.4      &            29.8            &              0               &              0              \\
 \ce{HfSe2}(T) &    0.33     &     -1.31     &            4.1             &              0               &              0              \\
               &    0.67     &     -1.22     &            58.6            &              0               &              0              \\
               &      1      &     -1.14     &           111.7            &              0               &              0              \\
\rowcolor{Gray}
 \ce{HfTe2}(T) &    0.33     &     -0.8      &            -2.1            &              0               &              0              \\
\rowcolor{Gray}
               &    0.67     &     -0.73     &            45.7            &              0               &              0              \\
\rowcolor{Gray}
               &      1      &     -0.66     &            95.7            &              0               &              0              \\
   \ce{InSe}   &    0.33     &     -0.49     &             24             &              0               &              0              \\
               &    0.67     &     -0.44     &            51.9            &              0               &              0              \\
               &      1      &     -0.42     &            52.8            &              0               &              0              \\
\rowcolor{Gray}
 \ce{MoS2}(H)  &    0.33     &     -0.75     &           118.3            &              0               &              0              \\
\rowcolor{Gray}
               &    0.67     &     -0.68     &           146.9            &              0               &              0              \\
\rowcolor{Gray}
               &      1      &     -0.62     &           169.1            &              0               &              0              \\
 \ce{MoSe2}(H) &    0.33     &     -0.51     &           158.3            &              0               &              0              \\
\rowcolor{Gray}
 \ce{MoTe2}(H) &    0.33     &     -0.11     &           189.9            &             0.02             &              0              \\
 \ce{NbS2}(H)  &    0.33     &     -1.17     &           -51.6            &              0               &              0              \\
               &    0.67     &     -1.17     &           -12.1            &              0               &              0              \\
               &      1      &     -1.13     &            55.5            &              0               &              0              \\
\rowcolor{Gray}
 \ce{NbS2}(T)  &    0.33     &     -1.1      &            23.6            &              0               &              0              \\
\rowcolor{Gray}
               &    0.67     &     -1.05     &            103             &              0               &              0              \\
 \ce{NbSe2}(H) &    0.33     &     -0.9      &           -10.3            &              0               &              0              \\
               &    0.67     &     -0.88     &            22.7            &              0               &              0              \\
               &      1      &     -0.82     &            96.4            &              0               &              0              \\
\hline
\end{tabular}}
    \caption{
        List of ic-2D materials with energy above the convex hull, $\Delta H_{\mathrm{hull}} < 200$ meV/atom. For each material, we report the chemical formula of the
        parent monolayer, the intercalated fraction, the heat of formation, the energy above the convex hull, 
        the band gap, and the magnetic moment.
    }
    \label{tab:TS1}
\end{table}

\begin{table}[ht]
    \centering
    \resizebox{0.67\textwidth}{!}{\begin{tabular}{lccccc}
\hline \hline
                Parent ML  &  ic-fraction  &  $\Delta H$ (eV/atom)  &  $\Delta H_{\mathrm{hull}}$ (meV/atom)  &  $E_{\mathrm{gap}}$ (eV) & $\mu_{\mathrm{mag}}$  \\
\hline
\rowcolor{Gray}
 \ce{NbSe2}(T) &    0.33     &     -0.85     &            41.8            &              0               &              0              \\
\rowcolor{Gray}
               &    0.67     &     -0.8      &           103.8            &              0               &              0              \\
\rowcolor{Gray}
               &      1      &     -0.75     &           161.9            &              0               &              0              \\
 \ce{NbTe2}(T) &    0.33     &     -0.4      &            68.8            &              0               &              0              \\
               &    0.67     &     -0.36     &           116.4            &              0               &              0              \\
               &      1      &     -0.31     &            176             &              0               &              0              \\
\rowcolor{Gray}
   \ce{PdS2}   &    0.25     &     -0.27     &            30.5            &             0.75             &              0              \\
\rowcolor{Gray}
               &     0.5     &     -0.26     &            53.4            &             0.6              &              0              \\
\rowcolor{Gray}
               &    0.75     &     -0.24     &            77.1            &             0.66             &              0              \\
  \ce{PdSe2}   &    0.25     &     -0.28     &            13.1            &             0.79             &              0              \\
               &     0.5     &     -0.27     &            32.8            &             0.71             &              0              \\
               &    0.75     &     -0.25     &            51.2            &             0.65             &              0              \\
\rowcolor{Gray}
 \ce{SnS2}(T)  &    0.33     &     -0.38     &            35.5            &              0               &              0              \\
\rowcolor{Gray}
               &    0.67     &     -0.36     &            62.1            &             0.02             &              0              \\
\rowcolor{Gray}
               &      1      &     -0.37     &            64.2            &              0               &              0              \\
 \ce{SnSe2}(T) &    0.33     &     -0.36     &            9.8             &              0               &              0              \\
               &    0.67     &     -0.36     &            17.5            &              0               &              0              \\
               &      1      &     -0.38     &            16.8            &              0               &              0              \\
\rowcolor{Gray}
   \ce{SnTe}   &    0.25     &     -0.18     &            87.1            &              0               &              0              \\
\rowcolor{Gray}
               &     0.5     &     -0.14     &            117             &              0               &              0              \\
 \ce{TaS2}(H)  &    0.33     &     -1.13     &           -75.3            &              0               &              0              \\
               &    0.67     &     -1.09     &           -70.2            &              0               &              0              \\
               &      1      &     -1.01     &           -17.8            &              0               &              0              \\
\rowcolor{Gray}
 \ce{TaS2}(T)  &    0.33     &     -1.05     &            2.3             &              0               &              0              \\
\rowcolor{Gray}
               &    0.67     &     -0.99     &            31.4            &              0               &              0              \\
\rowcolor{Gray}
               &      1      &     -0.91     &            73.1            &              0               &              0              \\
 \ce{TaSe2}(T) &    0.33     &     -0.77     &            34.5            &              0               &              0              \\
               &    0.67     &     -0.66     &           105.4            &              0               &              0              \\
               &      1      &     -0.56     &           175.6            &              0               &              0              \\
\rowcolor{Gray}
 \ce{TiTe2}(T) &    0.33     &     -0.72     &            -6.1            &              0               &              0              \\
\rowcolor{Gray}
               &    0.67     &     -0.69     &            28.7            &              0               &              0              \\
\rowcolor{Gray}
               &      1      &     -0.65     &            74.1            &              0               &            4.97             \\
  \ce{VS2}(H)  &    0.33     &     -0.96     &            -26             &              0               &            1.51             \\
               &    0.67     &     -0.96     &            23.8            &              0               &              0              \\
\rowcolor{Gray}
   \ce{W2N3}   &    0.33     &     -0.32     &            -135            &              0               &              0              \\
\rowcolor{Gray}
               &    0.67     &     -0.31     &           -121.1           &              0               &              0              \\
\rowcolor{Gray}
               &      1      &     -0.2      &           -15.6            &              0               &              0              \\
 \ce{WSe2}(T') &    0.33     &     -0.41     &           172.7            &              0               &             2.1             \\
               &    0.25     &     0.03      &           165.8            &              0               &              0              \\
               &     0.5     &     0.02      &           146.8            &              0               &            2.46             \\
\rowcolor{Gray}
   \ce{ZrBr}   &    0.33     &     -0.81     &           187.2            &              0               &              0              \\
  \ce{ZrCl2}   &    0.25     &     -1.48     &           165.3            &              0               &            2.08             \\
               &    0.33     &     -1.14     &           135.4            &              0               &             2.2             \\
               &    0.67     &     -1.03     &           198.7            &              0               &              0              \\
\rowcolor{Gray}
 \ce{ZrSe2}(T) &    0.33     &     -1.38     &           -60.4            &              0               &              0              \\
\rowcolor{Gray}
               &    0.67     &     -1.35     &           -65.7            &              0               &              0              \\
\rowcolor{Gray}
               &      1      &     -1.3      &           -49.8            &              0               &              0              \\
\hline
\end{tabular}}
    \caption{
        continuation of Table \ref{tab:TS1}
    }
\end{table}

\bibliography{references}